Review paper

# Magnetocaloric effects in RTX intermetallic compounds (R = Gd－Tm, T = Fe－Cu and Pd, X = Al and Si)

Zhang Hu(张虎)[a) b)] and Shen Bao-Gen(沈保根)[b)]

[a)]School of Materials Science and Engineering, University of Science and Technology of Beijing, Beijing 100083, China

[b)]State Key Laboratory for Magnetism, Institute of Physics, Chinese Academy of Sciences, Beijing 100190, China

**Abstract**

The ternary intermetallic RTX compounds (R = rare earth, T = transitional metal, X = *p*-block metal) have been investigated extensively in the past few decades due to their interesting physical properties. Recently, much attention has been paid to the magnetocaloric effect (MCE) of these RTX compounds, especially the ones with heavy rare-earth, for their potential application in low temperature magnetic refrigeration. In this paper, we review the MCE of RTSi and RTAl systems with R = Gd－Tm, T = Fe－Cu and Pd, which are widely investigated in recent years. It is found that these RTX compounds exhibit various crystal structures and magnetic properties, which then result in different MCE. Large MCE has been observed not only in the typical ferromagnetic materials but also in the antiferromagnetic materials. The magnetic properties have been studied in detail to discuss the physical mechanism of large MCE in RTX compounds. Particularly, some RTX compounds, such as ErFeSi, HoCuSi, HoCuAl, etc, exhibit large reversible MCE under low magnetic field change, which suggests that these compounds could be promising materials for magnetic refrigeration in low temperature range.

Project supported by the National Natural Science Foundation of China (Grant Nos. 51371026, 11274357 and 51327806) and the Fundamental Research Funds for the Central Universities (FRF-TP-14-011A2 and FRF-TP-15-002A3).

Corresponding author. E-mail: zhanghu@ustb.edu.cn and shenbg@aphy.iphy.ac.cn

# 1. Introduction

Nowadays, magnetic materials have been widely used and impact almost every aspect in our society from household appliances to aerospace sciences. Especially, the functional magnetic materials, such as permanent magnets, soft magnets, and magnetic shape memory alloys, etc, have played an essential role in the development of modern society. In recent years, magnetic refrigeration based on magnetocaloric effect (MCE) has been demonstrated to be a novel application of functional magnetic materials. Compared with conventional gas compression-expansion refrigeration, magnetic refrigeration technique has attracted considerable attention due to its great advantages in many aspects such as energy saving and environmental friendly.[1-3] As the core part of magnetic refrigeration technique, the magnetocaloric properties of magnetic materials greatly affect the performance of magnetic refrigerator, and thus, it is of importance to develop magnetic refrigerants with large MCE.

Since the discovery of giant MCE in $Gd_5(Si_{1-x}Ge_x)_4$,[4] a great deal of effort has been made to search suitable refrigerants for room temperature magnetic refrigeration.[5-13] On the other hand, it is also significant to search suitable materials exhibiting large MCE at low temperature due to their potential applications in gas liquefaction and scientific research.[2, 14] Usually, the magnitude of MCE can be characterized by magnetic entropy change ($\Delta S_M$) and/or adiabatic temperature change ($\Delta T_{ad}$) upon the variation of magnetic field. Thermodynamic analysis reveals that the maximum magnetic entropy value ($S_M$) per mole of magnetic ions is equal to $S_M = R\ln(2J+1)$, where $R$ is the universal gas constant and $J$ is the total angular momentum of a magnetic ion.[1] Therefore, a large $\Delta S_M$ can be usually expected in the heavy rare earth-based materials due to the high magnetic moments of heavy rare-earth atoms.

In the past few years, many different rare earth (R) – transition metal (T) intermetallic systems have been reported to exhibit large MCE in a wide temperature range.[14-16] Among them, the ternary intermetallic RTX compounds (R = rare earth, T = transitional metal, X = $p$-block metal) have been studied extensively due to their interesting physical properties and large MCE. It has been found that the magnetic moments of RTX compounds are mainly contributed by the rare earth atoms while the T and X atoms hardly contribute the magnetic moments due to the hybridization between $d$ states of T and $p$ states of X atoms. However, the crystallographic structure would change with the variation of either T or X atoms, thus also affecting the magnetic properties and MCE of RTX compounds. Very recently, Gupta et al.[17] reviewed the magnetic and related properties of RTX compounds, and briefly discussed the MCE. In the present paper, we give a comprehensive overview of the studies on the MCE of RTX intermetallic compounds (R = Gd－Tm, T = Fe－Cu and Pd, X = Al and Si), and this would be highly beneficial for the future research on the MCE of RTX compounds.

## 2. MCE in RTSi compounds

Table 1 lists the nature of magnetic ground state, the ordering temperature $T_{ord}$, and the magnetocaloric properties for RTSi (R = Gd－Er, T = Fe－Cu) compounds. It is seen that the RFeSi and RCoSi compounds order ferromagnetically except HoFeSi, while the RNiSi and RCuSi compounds exhibit antiferromagnetic (AFM) ground state. The ordering temperature $T_{ord}$ decreases with the rare earth atom sweeping from Gd to Er. In contrast, the MCE increases largely with the variation of rare earth atom from Gd to Er. It is interesting to note that large MCE can be observed not only in ferromagnetic (FM) RTSi but also in AFM RTSi compounds, which show field-induced metamagneic transition from AFM to FM states.

In the following sections, the magnetocaloric properties of RTSi compounds, especially the ones with large MCE, will be discussed in detail.

## 2.1 RFeSi compounds

The crystal structure of CeFeSi and related compounds was first investigated by Bodak *et al*. in 1970,[18] and they found that these compounds crystallize in the tetragonal CeFeSi-type structure (space group *P*4/*nmm*). In 1992, Welter *et al*.[19] studied the magnetic properties of RFeSi (R = La–Sm, Gd–Dy) compounds by susceptibility measurements and neutron diffraction studies. It was found that the RFeSi (R = Gd, Tb, and Dy) exhibit FM state below their respective $T_C$. Later, Napoletano *et al*.[20] reported that GdFeSi undergoes a FM-paramagnetic (PM) transition around 118 K and presents a large $-\Delta S_M$ of 22.3 J/kg K and $\Delta T_{ad}$ of 4.5 K for a high field change of 9 T. Especially, it shows a giant refrigerant capacity (RC) value of 1940 J/kg in the 10-160 K temperature range for a field change of 9 T. Recently, Zhang *et al*.[14, 21, 22] investigated the magnetic properties and MCE of RFeSi (R = Gd–Er) compounds systematically. The maximum $-\Delta S_M$ values of GdFeSi are 6.0 and 11.3 J/kg K for the field changes of 2 and 5 T, respectively. In addition, the RC value, calculated by integrating numerically the area under the $\Delta S_M$–$T$ curve with defining the temperatures at half maximum of peak as the integration limits,[23] is obtained to be 373 J/kg for a field change of 5 T. Figure 1 shows the temperature (*T*) dependence of zero-field-cooling (ZFC) and field-cooling (FC) magnetizations (*M*) under 0.05 T for RFeSi (R = Gd-Er) compounds.[14, 21, 22] An obvious difference between ZFC and FC curves appears below the ordering temperature for RFeSi compounds except GdFeSi, which may be due to the domain-wall-pinning effect as usually observed in materials with low ordering temperature and high anisotropy.[14, 24] The RFeSi (R = Tb and Dy) compounds experience a second-order FM-PM transition around the respective $T_C$ of 110 and 70 K for TbFeSi and DyFeSi, which are quite close to the liquefaction temperatures of natural gas (111 K) and nitrogen (77 K).[21] In addition, an unusual discrepancy between ZFC and FC curves can be observed in PM state of RFeSi (R = Tb and Dy) compounds, suggesting the existence of short-range FM correlations just above $T_C$.[25] The magnetic entropy change $\Delta S_M$ of RFeSi (R = Tb and Dy) was calculated from the magnetization isotherms by using Maxwell relation $\Delta S_M(T,H) = \mu_0 \int_0^H (\partial M / \partial T)_H dH$,[26] and the temperature dependence of $\Delta S_M$ for TbFeSi and DyFeSi under different magnetic field changes are shown in Fig 2(a).[21] It can be seen that TbFeSi and DyFeSi present large $-\Delta S_M$ values of 5.3 and 4.8 J/kg K for a low field change of 1 T, respectively. This large MCE under a low field change is favorable to practical applications since the maximum field of permanent magnets in market is usually lower than 2 T. In addition, it is worth noting that the magnitude of MCE is nearly same for TbFeSi and DyFeSi. Therefore, a series of $(Tb_{1-x}Dy_x)$FeSi compounds can be predicted theoretically to exhibit continuous ordering temperatures with similar magnitude of MCE. Figure 2(b) shows the temperature dependence of calculated $\Delta S_M$ of $(Tb_{1-x}Dy_x)$FeSi compounds for a field change of 1 T.[21] Furthermore, a composite material can be formed based on this series of $(Tb_{1-x}Dy_x)$FeSi compounds and the optimum mass ratio $y_i$ of each component, determined by using a numerical method,[27] is as follows: $y_1$ = 19.43 wt.%, $y_2$ = 13.32 wt.%, $y_3$ = 13.47 wt.%, $y_4$ = 13.74 wt.%, $y_5$ = 15.08 wt.%, and $y_6$ = 24.96 wt.% for $x$ = 0, 0.2, 0.4, 0.6, 0.8, and 1.0, respectively. The $\Delta S_M$ of this composite is estimated by using the equation $\Delta S_{com} = \sum_{i=1}^{6} y_i \Delta S_i$ and is shown in Fig. 2(b).[21] The composite exhibits a constant $-\Delta S_{com}$ of ~1.4 J/kg K in the wide temperature range, thus resulting in a large RC of 64 J/kg for a field change of 1 T, which is 49 % and 64 % higher than those of TbFeSi (43 J/kg) and DyFeSi (39 J/kg). Thermodynamic analysis indicates that a magnetic refrigeration system based on an ideal Ericsson cycle requires constant $\Delta S_M$ over a wide temperature range.[1] Therefore, above result suggests that the composite of $(Tb_{1-x}Dy_x)FeSi$ can be a good candidate of magnetic refrigerants for Ericsson cycle over the liquefaction temperatures of nitrogen and natural gas. Large reversible MCE for relatively low magnetic field change has also been observed in ErFeSi compound.[14] Figure 3 shows the temperature dependence of $\Delta S_M$ and $\Delta T_{ad}$ of ErFeSi for different magnetic field changes.[14] Here, the $\Delta T_{ad}$ was calculated from the heat capacity ($C_P$) curves by using the equation $\Delta T_{ad}(\Delta H, T) = [T(S)_H - T(S)_0]_S$.[26] For a magnetic field change of 5 T, the maximum values of $-\Delta S_M$ and $\Delta T_{ad}$ around the $T_C$ = 22 K are 23.1 J/kg K and 5.7 K, respectively. Particularly, the $-\Delta S_M$ and RC reach as high as 14.2 J/kg K and 130 J/kg for a relatively low field change of 2 T, respectively. This large MCE around liquid hydrogen temperature (20.3 K) indicates that ErFeSi could be promising material for magnetic refrigeration of hydrogen liquefaction.

Unlike other RFeSi compounds with FM ground state, HoFeSi exhibits a complex magnetic structure with FM and AFM/ferrimagnetic (FIM) moments at low temperatures.

With the decrease of temperature, HoFeSi undergoes a PM-FM transition at $T_C$ = 29 K. Besides, another anomaly is found at $T_t$ = 20 K, which indicates that part of magnetic moments in HoFeSi may experience a FM-AFM/FIM transition around $T_t$.[22] Figure 4(a) shows the magnetization isotherms of HoFeSi in the temperature range of 8-24 K.[22] It is seen that the magnetization below 20 K increases largely at first with increasing magnetic field, corresponding to the typical FM behavior. A field-induced metamagnetic transition occurs at critical field with further increase of field, suggesting the possible presence of AFM or FIM components at low temperatures in HoFeSi compound. The fraction of FM components, estimated by extrapolating the plateau of FM state to 5 T, is about 79% at 8 K and reaches nearly 100% when temperature increases to 24 K (see inset of Fig. 4(a)). The temperature dependence of $\Delta S_M$ for HoFeSi under different magnetic field changes is shown in Fig. 4(b).[22] It is noted that HoFeSi presents negative $\Delta S_M$ peak (normal MCE) around $T_C$ as well as positive $\Delta S_M$ (inverse MCE) around $T_t$. For a relatively low field change of 2 T, the $\Delta S_M$ values are 5.6 J/kg K at $T_t$ and −7.1 J/kg K at $T_C$, respectively. This special feature of successive inverse and normal MCE in HoFeSi could be applied in some refrigerators with special designs and functions, which other materials with only normal MCE cannot satisfy.[28]

## 2.2 RCoSi compounds

The crystal structures and magnetic properties of RCoSi compounds vary with the rare earth elements. The neutron diffraction studies revealed that RCoSi (R = Gd and Tb) crystallize in the tetragonal structure of CeFeSi-type (space group $P4/nmm$) and order antiferromagnetically below $T_N$ of 175 and 140 K for R = Gd and Tb, respectively.[29] However, RCoSi (R = Dy, Ho, and Er) compounds have been reported to crystallize in the orthorhombic TiNiSi-type crystal structure and exhibit PM-FM transition at low temperatures.[30-32] Leciejewicz et al.[31] investigated the magnetic structure of HoCoSi by neutron diffraction and found that the magnetic moments order ferromagnetically below $T_C$ = 13 K. Besides, the magnetic moments of Ho atoms form a conical spiral at 1.7 K, leading to the coexistence of the collinear FM structure and helicoidal structure.

In 2012, Xu et al.[32] further investigated the MCE of HoCoSi compound systematically by magnetization and heat capacity measurements. Figure 5 displays the heat capacity ($C_P$) curves for HoCoSi under different magnetic fields.[32] A distinct $\lambda$-type peak is observed around 11.2 K in zero field, corresponding to the second-order FM-PM transition. In addition, another anomaly is observed at $T_t$ = 4 K in the thermomagnetic curve (inset of Fig. 5), corresponding to the critical temperature of the coexistence of the collinear FM structure and helicoidal structure. With the increase of magnetic field, the peak gradually becomes broader and lower while it also slightly moves to higher temperature, suggesting the typical characteristic of ferromagnet.[1] Based on the theory of thermodynamics, the $\Delta S_M$ and $\Delta T_{ad}$ values can be calculated from the $C_P$ curves by using the following equations $\Delta S_M(T) = \int_0^T [C_H(T) - C_0(T)]/TdT$ and $\Delta T_{ad}(\Delta H,T) = [T(S)_H - T(S)_0]_S$, respectively. Figure 6 shows the temperature dependence of $\Delta S_M$ and $\Delta T_{ad}$ for HoCoSi under different magnetic field changes.[32] It is found that the $-\Delta S_M$ and $\Delta T_{ad}$ reach as high as 26.3 J/kg K and 11.0 K for a magnetic field change of 5 T. Moreover, for a relatively low field change of 2 T, HoCoSi compound exhibits a giant MCE around $T_C$ = 15 K with the maximum $-\Delta S_M$ of 17.3 J/kg K and $\Delta T_{ad}$ of 6.2 K. Very recently, Gupta et al.[33] also reported the MCE of HoCoSi compound and observed a large MCE without hysteresis loss around the $T_C$ of 14 K. For a magnetic field change of 5 T, the maximum $-\Delta S_M$ and RC values of HoCoSi are 20.5 J/kg K and 410 J/kg, respectively.

Xu et al.[32] further studied the effect of Er substitution on the magnetic and magnetocaloric properties in $(Ho_{1-x}Er_x)CoSi$ compounds. Figure 7 shows the temperature dependence of ZFC and FC magnetizations under 0.01 T for $(Ho_{1-x}Er_x)CoSi$ compounds.[32] In addition to the FM-PM transition around $T_C$, all $(Ho_{1-x}Er_x)CoSi$ compounds except ErCoSi exhibit another anomaly at lower temperature $T_t$, which is related to the presence of magnetic helicoidal structure. It is clearly seen that the transition temperatures decrease linearly with the Er content increasing from 0 to 1 (inset of Fig. 7). Figure 8 shows the temperature dependence of $\Delta S_M$ for $(Ho_{1-x}Er_x)CoSi$ compounds under a field change of 2 T.[32] For a field change of 2 T, the maximum $-\Delta S_M$ values are 17.9, 17.9, 18.2, 18.0, 18.5, and 18.7 J/kg K for $x$ = 0, 0.2, 0.4, 0.6, 0.8, and 1, respectively. It can be seen that $(Ho_{1-x}Er_x)CoSi$ compounds exhibit nearly same magnitude of MCE with increasing the Er content. Meanwhile, the $T_C$ decreases from 15 to 5.5 K with the variation of $x$ from 0 to 1. Therefore, a composite material can be constructed based on this series of $(Ho_{1-x}Er_x)CoSi$ compounds and exhibits a constant $\Delta S_M$ in the temperature range of 5.5-15 K, satisfying the requirement of Ericsson-cycle magnetic refrigeration. Similar research has also been carried out in $(Ho_{1-x}Dy_x)CoSi$ compounds.[32] However, single phase with tetragonal CeFeSi-type structure can not be obtained when Dy content is higher than 0.4 due to the presence of impurity $DyCo_2Si_2$ phase. It is found that the $T_C$ decreases from 15 K for $x$ = 0 to 5 K for $x$ = 0.4 in $(Ho_{1-x}Dy_x)CoSi$ compounds. Figure 9 shows the temperature dependence of $\Delta S_M$ for Ho$_{0.8}$Dy$_{0.2}$CoSi compound under different magnetic field changes.[32] For a magnetic field change of 5 T, Ho$_{0.8}$Dy$_{0.2}$CoSi exhibits a maximum $-\Delta S_M$ value of 20.2 J/kg K around $T_C$ = 12 K. The reduction of MCE is due to the fact that the introduction of Dy atoms would result in the competition of FM coupling of Ho moments and AFM coupling of Dy moments, and thus lowering the saturation magnetization and MCE of (Ho$_{1-x}$Dy$_x$)CoSi compounds.

## 2.3 RNiSi compounds

In 1974, Bodak *et al.*[34] reported that all RNiSi (R = Gd-Lu) compounds crystallize in the orthorhombic TiNiSi-type crystal structure. In 1999, Szytula *et al.*[35] further investigated the magnetic properties of RNiSi (R = Tb-Er) compounds by neutron diffraction and magnetometric measurements studies. It was found that all RNiSi (R = Tb-Er) compounds show AFM ordering with strong magnetocrystalline anisotropy at low temperatures. In addition, the neutron diffraction studies revealed that another ordering change of magnetic moments from sine to square-modulated structure occurs below $T_N$. Very recently, the MCE of RNiSi (R = Tb-Er) compounds have been investigated by different researchers.[36-38] Among these materials, HoNiSi exhibits the largest MCE due to the field-induced metamagnetic transition.[36]

Zhang *et al.*[37] recently reported the giant rotating MCE in textured DyNiSi polycrystalline material, which is larger than those of most rotating magnetic refrigerants reported so far. Figure 10 shows the temperature dependence of ZFC and FC magnetizations for DyNiSi at 0.05 T along the parallel and perpendicular directions, respectively.[37] It is seen that both thermomagnetic curves show similar trend but with different magnetizations. With the decrease of temperature, DyNiSi experiences a PM-AFM transition at $T_N$ of 8.8 K. In addition, another anomaly is found around the transition temperature $T_t$ = 4 K, which is likely attributed to the ordering change of magnetic moments from sine to square-modulated structure based on the neutron diffraction studies.[35] The magnetization was measured by rotating the DyNiSi sample in the magnetic field of 0.05 T as shown in the inset of Fig. 10. Here, the rotation angle $\theta$ is defined as 0° when the longitudinal direction of columnar grains is parallel to the magnetic field. It can be clearly seen that the magnetization decreases gradually by rotating the sample from parallel to perpendicular direction, indicating that the easy magnetization axis is consistent with the preferred crystalline orientation.

Figures 11(a) and (b) show the temperature dependence of $\Delta S$ for DyNiSi for different magnetic field changes along parallel and perpendicular directions, respectively.[37] It can be seen that DyNiSi exhibits a giant anisotropic MCE, e.g., the maximum $-\Delta S$ values are 22.9 and 5.9 J/kg K for a field change of 5 T along parallel and perpendicular directions, respectively. A small negative $-\Delta S$ value (inverse MCE) is observed below $T_N$ at 1 T along parallel direction because of the presence of AFM state, while it becomes positive with the increase of magnetic field. Such a sign change of $-\Delta S$ is due to the field-induced AFM-FM metamagnetic transition.[39] In addition, another $-\Delta S$ peak is found around $T_t$ for both directions. It is speculated that the transition from sine to square-modulated structure may lead to some unstable moments below $T_t$. Therefore, an applied magnetic field will turn these AFM components into FM ordering which exhibits a magnetically more ordered configuration, and then resulting in a positive $-\Delta S$ peak. Figure 11(c) shows the difference of $\Delta S$ for DyNiSi between different directions as a function of temperature for different magnetic field changes.[37] For the field changes of 2 and 5 T, the $-\Delta S_{diff}$ peak reaches as high as 11.1 J/kg K at 8.5 K and 17.6 J/kg K at 13 K due to the giant anisotropy of MCE. This result indicates that a large rotating MCE can be obtained by rotating the sample from perpendicular to parallel direction.

Furthermore, the isothermal magnetization curves at 8 and 9 K were measured for DyNiSi under applied fields up to 2 T by rotating the sample from perpendicular (90°) to parallel (0°) direction with a step of 10° as shown in inset of Fig. 12.[37] By defining the rotating entropy change $\Delta S^R(90°)$ as zero, the $\Delta S^R(\theta)$ value at 8.5 K can be obtained based on the magnetization curves by using the following equation:

$$\Delta S^R(\theta) = \Delta S(\theta) - \Delta S(90°)$$
$$= \mu_0 \int_0^H (\frac{\partial M(\theta)}{\partial T})_H dH - \mu_0 \int_0^H (\frac{\partial M(90°)}{\partial T})_H dH \quad (1)$$

Figure 12 shows the $\Delta S^R(\theta)$ as a function of rotation angle for different magnetic field changes.[37] It is noted that small negative $-\Delta S^R(\theta)$ value is observed under relatively low fields near the perpendicular direction. This can be understood that the disordering of magnetic moments in AFM sublattice is enhanced under low fields when the rotation just starts, and thus it leads to the positive entropy change. With further rotating the sample towards parallel direction, the majority of spins in AFM sublattice could orient along the field direction, which then increases the spin ordering and results in the positive $-\Delta S^R(\theta)$ value. Under the magnetic field of 2 T, the $-\Delta S^R(\theta)$ value increases gradually and reaches a maximum value of 7.9 J/kg K as the sample is rotated from perpendicular to parallel direction.

Figures 13(a) and (b) show the temperature dependence of ZFC and FC magnetizations for HoNiSi[36] and ErNiSi under 0.05 T. It is seen that HoNiSi and ErNiSi undergo an AFM-PM transition around the Néel temperature $T_N$ = 3.8 K and 4 K, respectively, which are just around the critical temperature of liquid helium (4 K), suggesting the potential application of RNiSi (R = Ho and Er) for helium liquefaction. Besides, no obvious discrepancy between ZFC and FC curves is observed, indicating the good thermomagnetic reversibility of the magnetic transition. The temperature dependence of magnetization in various magnetic fields is shown in the inset of Figs. 13(a) and (b). It can be seen that RNiSi (R = Ho and Er) exhibit typical AFM-PM transition under low fields. However, the magnetization at low temperatures increases largely with increasing field, revealing the occurrence of a field-induced AFM-FM metamagnetic transition below $T_N$. In addition, a step-like behavior of $M$-$T$ curves above $T_N$ is also observed when field is larger than 0.3 T for HoNiSi and 0.7 T for ErNiSi, respectively, corresponding to the FM-PM transition.[40] Figures 14(a) and (b) show the temperature dependence of $\Delta S_M$ for HoNiSi[36] and ErNiSi under different magnetic field changes up to 5 T. It is well known that the $\Delta S_M$ values can be calculated either from the magnetization isotherms by using the Maxwell relation $\Delta S_M(T,H) = \mu_0 \int_0^H (\partial M/\partial T)_H dH$ or from the heat capacity by using the equation $\Delta S_M(T) = \int_0^T [C_H(T) - C_0(T)]/T dT$,[26] respectively. However, sometimes the values of $\Delta S_M$ calculated from heat capacity may be much lower than those obtained from magnetization isotherms, which is likely due to either false calculation of Maxwell relation in the vicinity of FOPT or the poor contact between sample and measuring platform during heat capacity measurement.[14, 41] For comparison, the $\Delta S_M$ values were estimated from both methods as shown in Fig. 14(a), and it can be clearly seen that the $\Delta S_M$ curves obtained from two methods match well with each other. For a field change of 2 T, the maximum $-\Delta S_M$ values for HoNiSi and ErNiSi around $T_N$ are 17.5 and 8.8 J/kg K, respectively. Besides, a large positive $\Delta S_M$ (inverse MCE) can be observed below $T_N$, which is caused by the presence of AFM ordering at low temperatures. For example, the $\Delta S_M$ of HoNiSi below $T_N$ reaches 7.2 J/kg K for a low field change of 0.5 T, and the maximum $\Delta S_M$ of ErNiSi below $T_N$ is 6.5 J/kg K for a low field change of 1 T. These large normal and inverse MCE under low field change indicate that RNiSi (R = Ho and Er) could be applied in magnetic refrigeration with either adiabatic magnetization or adiabatic demagnetization.

Recently, Franco *et al.*[42, 43] proposed a phenomenological procedure to construct the universal curve of $\Delta S_M$ for materials with second-order FM-PM transition. However, the applicability of this universal curve has not been proved in AFM materials. As shown in the inset of Fig. 14(a), Zhang *et al.*[36] first constructed the universal curve of $\Delta S_M$ for HoNiSi by using this phenomenological procedure. The normalized $\Delta S'$ is defined as $\Delta S'(T, H_{max}) = \Delta S_M(T, H_{max}) / \Delta S_M^{pk}(H_{max})$. The temperature axis has been rescaled in a different way below and above $T_{pk}$, by imposing that the positions of two reference points in the curve correspond to $\theta = \pm 1$,

$$\theta = \begin{cases} -(T - T_{pk})/(T_{r1} - T_{pk}), & T \leq T_{pk} \\ (T - T_{pk})/(T_{r2} - T_{pk}), & T > T_{pk} \end{cases} \quad (1)$$

Where $T_{r1}$ and $T_{r2}$ are the temperatures of the two reference points which correspond to $\frac{1}{2}\Delta S_M^{pk}$. All the curves under different field changes collapse onto the same universal curve though the ground state changes from AFM to FM with the increase of magnetic field. This result reveals that the universal $\Delta S_M$ curve could also be applied in AFM or at least weak AFM materials.

Figure 15 displays the temperature dependence of $\Delta T_{ad}$ for HoNiSi under different magnetic field changes.[36] HoNiSi exhibits large $\Delta T_{ad}$ values of 4.5 and 8.5 K for the field changes of 2 and 5 T, respectively, which is attributed to the field-induced metamagnetic transition from weak AFM to FM states.[15] In addition, it is found that both $\Delta S_M$ and $\Delta T_{ad}$ peaks for HoNiSi compound broaden asymmetrically towards high temperatures with increasing field, indicating the presence of FM ordering above $T_N$ induced by magnetic field.[44] Moreover, this broad distribution of $\Delta S_M$ peak of HoNiSi leads to a high RC value of 471 J/kg for a field change of 5 T.

## 2.4 RCuSi compounds

The series of ternary intermetallic RCuSi compounds have been investigated extensively in the past few decades due to the interesting physical properties. It has been reported that these compounds crystallize within two types of crystal structure depending on the annealing temperature. The high-temperature phase crystallizes in the $AlB_2$-type structure (space group $P6/mmm$) with R atoms at 1$a$: (0, 0, 0) and Cu/Si atoms statistically at 2$d$: (1/3, 2/3, 1/2).[45] The low-temperature phase adopts the $Ni_2In$-type structure (space group $P6_3/mmc$) with R at 2$a$: (0, 0, 0), Cu at 2$c$: (1/3, 2/3, 1/4), and Si at 2$d$: (1/3, 2/3, 3/4), respectively.[46, 47] The magnetic properties of RCuSi vary greatly with the change of crystal structure and rare earth element. According to the magnetic susceptibility measurements, Kido *et al*.[48] suggested that the $AlB_2$-type RCuSi compounds with R = Ce, Nd order antiferromagnetically while those with R = Gd, Ho order ferromagnetically. As for the RCuSi compounds with $Ni_2In$-type structure, neutron diffraction studies revealed that RCuSi compounds with R = Tb, Dy, Ho, and Er orders antiferromagnetically below $T_N$ = 16 K, 11 K, 9 K, and 6.8 K respectively.[49-52]

In 2010, Chen et al.[15, 53, 54] investigated systematically the magnetic properties and MCE of RCuSi (R = Gd-Er) with $Ni_2In$-type structure. It was found that these compounds show weak AFM ground state at low temperatures, which could be easily induced into FM state by magnetic field, and thus leading to large MCE. Figure 16 shows the temperature dependence of ZFC and FC magnetizations under the magnetic field of 0.01 T for RCuSi (R = Gd, Tb, Dy, and Er) compounds.[54] It can be found that the RCuSi compounds undergo an AFM-PM transition at $T_N$ = 14, 11, 10, and 7 K for R = Gd, Tb, Dy, and Er, respectively. The ZFC and FC curves are completely reversible above $T_N$, suggesting the good thermomagnetic reversibility of the magnetic transition. However, a small discrepancy is observed below the $T_N$, which is likely related to the domain-wall-pinning effect. In ZFC mode, the domain walls are pinned and the thermal energy is not strong enough to overcome the energy barriers, and this leads to the low magnetization at low temperatures. However, in FC mode, the magnetic field during the cooling prevents the pinning effect and therefore, the magnetization at low temperatures is higher than that in ZFC mode. Figure 17 shows the temperature dependence of $\Delta S_M$ for RCuSi (R = Gd, Tb, Dy, and Er) compounds under different magnetic field changes.[54] It is clearly seen that that RCuSi (R = Gd, and Tb) compounds exhibit a small negative $\Delta S_M$ value at a lower temperature, but the $\Delta S_M$ changes to positive value with increasing magnetic field, corresponding to the field-induced metamagnetic transition from AFM to FM states. This sign change of $\Delta S_M$ is not observed in DyCuSi and ErCuSi, indicating that the weak AFM coupling in RCuSi (R = Dy and Er) could be easily induced into FM state under lower field. Moreover, a large MCE can be obtained due to the field-induced metamagnetic transition. For a magnetic field change of 5 T, the maximum $-\Delta S_M$ values are 9.2, 10.0, 24.0, and 23.1 J/kg K for R = Gd, Tb, Dy, and Er, respectively.

Figure 18 shows the temperature dependence of magnetization for HoCuSi under various magnetic fields.[15] It is seen that the HoCuSi experiences an AFM to PM transition around $T_N$ of 7 K under low fields. With the increase of magnetic field, the magnetization at low temperatures increases largely, indicating the occurrence of a field-induced metamagnetic transition from AFM to FM states. In addition, a stepwise behavior of the M-T curves above $T_N$ is observed when the field is higher than 0.3 T, corresponding to the FM-PM transition. The $\Delta S_M$ of HoCuSi as a function of temperature for different magnetic field changes is shown in Fig. 19.[15] The maximum $-\Delta S_M$ and RC values are obtained to be 33.1 J/kg K and 385 J/kg for a magnetic field change of 5 T, respectively, which are comparable with or even higher than those of other refrigerant materials in similar temperature range. Particularly, the $-\Delta S_M$ reaches as high as 16.7 J/kg K for a relatively low field change of 2 T, making HoCuSi attractive candidate for magnetic refrigeration materials in the low temperature range.

The large MCE in HoCuSi compound is mainly attributed to the following reasons: (1) the high saturation magnetization ($M_S$ ~9.47 $\mu_B$),[15] (2) the field-induced metamagnetic transition from AFM to FM states,[55] and (3) the large change in lattice volume around $T_N$.[51] In order to investigate the change of lattice volume, the thermal expansion data ($\Delta L/L_{(50\ K)}$) under different fields has been measured by the means of strain gauge method[56] and is shown in Fig. 20.[54] The ($\Delta L/L_{(50\ K)}$) value decreases linearly with decreasing temperature above $T_N$ but drops abruptly around $T_N$ in zero magnetic field. This result confirms the occurrence of abrupt thermal expansion around $T_N$, which is caused by the change of lattice constants. In addition, it is noted that the abrupt thermal expansion shifts to higher temperature with the increase of field, and thus leading to the asymmetrical broadening of $-\Delta S_M$ peak.

## 3. MCE in RTAl compounds

The nature of magnetic ground state, the ordering temperature $T_{ord}$, and the magnetocaloric properties for RTAl (R = Gd－Tm, T = Fe－Cu and Pd) compounds are summarized in Table 2. Similar to the RTSi compounds with T = Fe, Co, Ni, RFeAl and RCoAl compounds order ferri-/ferro-magnetically, and RNiAl compounds exhibit AFM or AFM+FM ground state. Unlike RCuSi compounds exhibiting AFM ground state, RCuAl compounds order ferromagnetically at low temperatures. It is worth noting that RTAl (T = Fe and Pd) compounds could exist in various crystallographic structures depending on the different heat treatment techniques, and that would result in rich variety of magnetic properties and MCE. Among these materials, the largest MCE can be often observed in RTAl compounds with R = Ho due to the high value of total angular momentum $J$. In the following sections, the magnetocaloric properties of different series of RTAl compounds will be discussed in detail.

### 3.1 RFeAl compounds

It has been reported that all RFeAl (R = Gd-Dy) compounds crystallize in the hexagonal MgZn$_2$-type structure (space group $P6_3/mmc$) when quenched after annealing.[57-60] However, Klimczak *et al.*[57] found that GdFeAl crystallizes in cubic MgCu$_2$-type structure (space group $F$d3$m$) when cooled slowly, which exhibits lower saturation magnetic moments than that of GdFeAl with MgZn$_2$-type structure. In 1973, Oesterreicher et al.[61] reported that both GdFeAl and TbFeAl with MgZn$_2$-type structure order ferrimagnetically below the transition temperatures of 260 K and 195 K, respectively. In addition, a S-shape of *M-H* curve is observed in TbFeAl compound at low temperatures, which was ascribed to the partial chemical disorder of Fe and Al atoms as well as high magnetocrystalline anisotropy.[61, 62] Similar results have also been observed by Kastil et al.[59] In recent years, the MCE of RFeAl (R = Gd-Dy) have been reported by different researchers, and it is found that these compounds show reversible MCE in a wide temperature range, leading to a high RC value.[58-60]

In 2009, Dong et al.[58] first investigated the MCE of GdFeAl compound with MgZn$_2$-type structure. Figure 21(a) shows the temperature dependence of magnetization for GdFeAl under 0.1 T.[58] The transition temperature $T_C$ is 265 K, defined as the minimum value of d$M$/d$T$ curve. This $T_C$ is close to room temperature, indicating the possible application of GdFeAl compound for magnetic refrigeration near room temperature. The isothermal magnetization at 5 K is presented in Fig. 21(b).[58] The saturation magnetization $\mu_S$ is determined to be 5.8 $\mu_B$ per formula, which is lower than the theoretical g$J$ value of 7 $\mu_B$ for a free Gd$^{3+}$ ion. This lower value of $\mu_S$ is attributed to the AFM coupling between the magnetic moments of Gd and Fe sublattices.[63]

Figure 22 displays the temperature dependence of $\Delta S_M$ for GdFeAl under the magnetic field changes of 2 and 5 T, respectively.[58] The maximum $-\Delta S_M$ value is 3.7 J/kg K for a field change of 5 T, which is lower than those of most magnetic refrigerants in the same temperature range. However, the $\Delta S_M$ peak spreads out over a wide temperature range and the full width at half maximum of the peak is 159 K. This broad distribution of $\Delta S_M$ peak results in a high RC value of 420 J/kg for a field change of 5 T. In addition, a perfect magnetic reversibility around the transition temperature is observed in the *M-H* curves with the field increasing and decreasing modes, corresponding to the typical second-order magnetic transition. This result indicates that the detrimental effects for fast-cycling refrigerators of hysteresis losses and slow kinetics do not exist in GdFeAl compound. Very recently, Kastil *et al*.[59] also studied the MCE of RFeAl (R = Gd, Tb) and observed large relative cooling power (RCP) of 348 and 350 J/kg over a wide temperature region for GdFeAl and TbFeAl, respectively. Li *et al*.[60, 64] reported the MCE of RFeAl (R = Dy, Ho, and Er) and found that the RC of DyFeAl reaches the largest value of 832 J/kg for a field change of 7 T in this series of compounds.

### 3.2 RCoAl compounds

RCoAl compounds have been reported to crystallize in the hexagonal $MgZn_2$-type structure (space group *P*6$_3$/*mmc*) which is a closed-packed Laves phase.[65] In 2000, Jarosz *et al*.[66] investigated the crystallographic, electronic structure and magnetic properties of GdCoAl compound systematically, and reported that GdCoAl undergoes a typical FM-PM transition at $T_C$ = 100 K. Later, Zhang *et al*.[67] studied the magnetic entropy change of RCoAl (R = Gd - Ho) compounds in the temperature range of 10-100 K. Figure 23 presents the $T_C$ and $\Delta S_M$ for a field change of 5 T as a function of R atom. It is found that the $T_C$ decreases linearly with the rare earth atom varying from Gd to Tm. On the contrary, the $\Delta S_M$ has been reported to increase with the R atom changing from Gd to Tm. For a field change of 5 T, the HoCoAl exhibits the largest $-\Delta S_M$ of 21.5 J/kg K around $T_C$ = 10 K. In addition, a table-like $\Delta S_M$ peak was observed over the temperature range of 70 – 105 K in GdCoAl compound. Although the $-\Delta S_M$ of 10.4 J/kg K is relatively low for GdCoAl, this flat $\Delta S_M$ peak over a wide temperature range satisfies the requirement of a magnetic refrigerator based on an ideal Ericsson cycle, and also makes GdCoAl an attractive candidate material to fill the gap near 100 K in the $\Delta S_M$–$T$ profile required by an eight-stage magnetic refrigerator.[68] In 2010, Chelvane et al.[69] also investigated the magnetic and magnetocaloric properties of DyCoAl compound by magnetization and neutron diffraction measurements. It was found that DyCoAl has a collinear ferromagnetic structure where Dy moments lie in the ab plane at 10 K. A reversible MCE with –$\Delta S_M$ of 18 J/kg K for a field change of 9 T is obtained in DyCoAl around 37 K.

Very recently, Mo et al.[70] further studied the MCE of TmCoAl compound. Figure 24(a) shows the temperature dependence of magnetization under the magnetic field of 0.01 T for TmCoAl compound.[70] It can be found that the TmCoAl undergoes a typical second-order magnetic transition from FM to PM states around $T_C$ = 6 K, which is just above the boiling temperature of helium. A significant thermomagnetic irreversibility can be clearly seen below $T_C$, which is likely arising from the narrow domain-wall-pinning effect. The inverse dc susceptibility ($\chi^{-1}$) under 0.01 T and the Curie-Weiss fit to the experimental data for TmCoAl are plotted in the Fig. 24(b).[70] The inverse susceptibility above $T_C$ obeys the Curie-Weiss law $\chi^{-1} = (T - \theta_P)/C$, where $\theta_P$ is the paramagnetic Curie temperature and $C$ is the Curie-Weiss constant. Based on the calculation of Curie-Weiss fit, the values of $\theta_P$ and effective magnetic moment ($\mu_{eff}$) for TmCoAl compound are obtained to be 4 K and 5.93 $\mu_B$/Tm$^{3+}$, respectively. The $\mu_{eff}$ value is lower than the theoretical magnetic moment ($g\sqrt{J(J+1)}$ = 7.57 $\mu_B$) of Tm$^{3+}$ free ion, which is likely due to the crystal field effects and magnetic anisotropy. Figure 25 shows the temperature dependence of $\Delta S_M$ for TmCoAl under different magnetic field changes.[70] It is seen that TmCoAl exhibits a large $-\Delta S_M$ value of 10.2 J/kg K for a low magnetic field change of 2 T, comparable with or larger than those of most potential magnetic refrigerants with a similar magnetic transition temperature. Moreover, no thermal and magnetic hysteresis loss has been observed in TmCoAl compound. Therefore, the large reversible MCE suggests that TmCoAl could be a promising candidate for magnetic refrigeration at low temperatures.

### 3.3 RNiAl compounds

The RNiAl alloys have been intensively studied for their complex magnetic structures and related interesting physical properties.[66, 71-74] All RNiAl compounds crystallize in the ZrNiAl-type hexagonal structure (space group $P\bar{6}2m$). Neutron diffraction experiments revealed the coexistence of FM and AFM states in isostructural RNiAl compounds (R = Tb, Dy, and Ho) compounds.[71, 75, 76] Korte *et al*.[72] reported that GdNiAl compound experiences three transitions with FM ordering of the Gd spins at $T_C$ = 58 K accompanied by AFM processes at $T_1$ = 28 K and $T_2$ = 23 K, respectively. Similar result has also been observed by Si *et al*. in the study of annealed GdNiAl ribbon.[74] The successive magnetic transitions lead to a broad $\Delta S_M$ peak over a wide temperature range. By substituting Gd with Er, all transition temperatures shift to lower temperature while the MCE increases gradually. Singh *et al*.[77-79] investigated the magnetic and magnetocaloric properties of RNiAl (R = Tb, Dy, and Ho) in detail, and found that these compounds undergo two successive transitions with the decrease of temperature. For example, HoNiAl experiences a PM-FM transition at $T_C$ = 14 K followed by a FM-AFM transition at $T_1$ = 5 K, and exhibits a large $-\Delta S_M$ of 12.3 J/kg K around $T_C$ for a field change of 2 T.[79] Mo et al.[80] further reported that TmNiAl orders antiferromagnetically below 4 K, and shows a maximum $-\Delta S_M$ of 12.7 J/kg K for a field change of 5 T due to the metamagnetic transition from AFM to FM states.

In order to investigate the effect of Cu doping on the magnetic and magnetocaloric properties in the TmNiAl compound, Mo et al.[81] also studied the TmNi$_{1-x}$Cu$_x$Al compounds. Figure 26 displays the isothermal magnetization curves of TmNi$_{1-x}$Cu$_x$Al compounds as a function of magnetic field measured at 2 K in applied fields up to 5 T.[81] The magnetization increases linearly with increasing magnetic field in low-field ranges when $x < 0.3$, suggesting the existence of AFM ground state, and then exhibits a sharp increase at a critical field, confirming the field-induced metamagnetic transition from AFM to FM states. However, with increasing Cu-concentration in the $x \geq 0.3$, the magnetization increases rapidly with magnetic field and tends to be saturated at 5 T, which corresponds to the typical FM nature. The variation of ground state is attributed to the rotation of Tm magnetic moments from basal plane to $c$-axis, and thus leading to the canted AFM structure with larger projected moments along the $c$-axis near $T_{ord}$. Meanwhile, the $T_{ord}$ decreases from 4 K for $x = 0$ to 2.8 K for $x = 1$. Figure 27 shows the $\Delta S_M$ as a function of temperature for TmNi$_{1-x}$Cu$_x$Al compounds under a magnetic field change of 2 T.[81] It is found that the $-\Delta S_M$ value increases largely with the increase of Cu content, e.g., the $-\Delta S_M$ value of 10.7 J/kg K for TmNi$_{0.7}$Cu$_{0.3}$Al compound is almost twice that of TmNiAl compound (5.5 J/kg K). The MCE of TmNi$_{1-x}$Cu$_x$Al compounds with $x \geq 0.3$ are much higher than those of many magnetic refrigerant materials with a similar transition temperature.

In 2015, Cui et al.[82] reported the magnetic properties and MCE in HoNi$_{1-x}$Cu$_x$Al compounds. The temperature dependence of magnetization for HoNi$_{1-x}$Cu$_x$Al compounds under the magnetic field of 0.01 T is displayed in Fig. 28. The HoNi$_{1-x}$Cu$_x$Al compounds with $x \leq 0.1$ exhibit two magnetic transitions, which are speculated to be a PM-FM+AFM transition followed by an AFM-AFM transition. These complex magnetic transitions are induced by the combination and competition between FM and AFM orderings. With further increasing Cu content, the compounds with $x = 0.2$-$0.7$ undergo a single AFM-PM transition, and the ones with $x = 0.8$-$1$ are found to show a FM ground state at low temperatures. For comparison, Figure 29 shows the $\Delta S_M$ as a function of temperature for HoNi$_{1-x}$Cu$_x$Al with $x = 0.3$ and $0.8$ under different magnetic field changes.[82] A small negative value of $-\Delta S_M$ was observed for a low magnetic field change of 1 T in $x = 0.3$ compound, corresponding to the presence of AFM state at low temperatures. On the contrary, the compound with $x = 0.8$ exhibits positive $-\Delta S_M$ for all magnetic field changes, which is due to the typical FM-PM transition. Large $-\Delta S_M$ values of 12.3 and 9.4 J/kg K for a low field change of 2 T are obtained in HoNi$_{1-x}$Cu$_x$Al with $x = 0.3$ and $0.8$, respectively.

Wang et al.[83, 84] also reported similar work on ErNi$_{1-x}$Cu$_x$Al compounds. Figure 30 shows the temperature dependence of magnetization for ErNi$_{1-x}$Cu$_x$Al compounds with $x = 0.2$, 0.5, and 0.8, respectively.[83] It is found that the sample with $x = 0.2$ orders antiferromagnetically below the $T_N = 4.6$ K, while the one with $x = 0.5$ orders ferromagnetically around the $T_C = 5.8$ K. A quasi Curie-like magnetic transition is observed at $T_{trs} = 5.5$ K for $x = 0.8$ sample. In order to further investigate the nature of this magnetic transition, the ac susceptibilities in different frequencies for the samples with $x = 0.5$ and 0.8 were measured and plotted in Fig. 31.[83] It is clearly seen that the peak position of ac susceptibilities for $x = 0.5$ is independent of frequency. On the contrary, the peak for $x = 0.8$ shifts toward higher temperatures with increasing the frequency as shown in the inset of Fig. 31(b), which suggests the possible existence of short-range order. Figure 32 shows temperature dependence of $-\Delta S_M$ under different magnetic field changes for ErNi$_{1-x}$Cu$_x$Al with $x$ = 0.2, 0.5, and 0.8, respectively.[83] The sample with $x = 0.2$ presents a small negative value of $-\Delta S_M$ below $T_N$ under a low magnetic field change of 1 T, corresponding to the nature of AFM state at low temperatures. With the increase of magnetic field, the field-induced metamagnetic transition leads to a large positive $-\Delta S_M$ for compound with $x = 0.2$. For a relatively low field change of 2 T, the maximum $-\Delta S_M$ values are 10.1, 14.7, and 15.7 J/kg K for $x$ = 0.2, 0.5, and 0.8, respectively, and this large MCE under low field change is favorable for practical applications.

### 3.4 RCuAl compounds

The RCuAl compounds, like RNiAl compounds, also crystallize in the ZrNiAl-type hexagonal structure (space group $P\bar{6}2m$).[66, 85, 86] However, unlike RNiAl compounds, all heavy rare-earth RCuAl compounds exhibit a FM ground state.[87] There are two types of basal plane layers distributed along the $c$-axis: one contains all the R atoms and one-third of Cu atoms, and the other contains a nonmagnetic layer formed by all the Al atoms and two-third of Cu atoms. This layered character of the crystalline structure leads to large uniaxial magnetic anisotropies in GdCuAl, DyCuAl and ErCuAl, and a basal-plane type of magnetic anisotropy in HoCuAl.[88]

Recently, Dong et al.[85, 86, 89-91] studied the MCE of RCuAl compounds systematically. Figure 33 shows the temperature dependence of ZFC and FC magnetizations for crystalline RCuAl (R = Gd-Er) compounds under 0.1 and 0.05 T. These compounds undergo a typical FM-PM transition, and the $T_C$ decreases monotonically with the R atom varying from Gd to Er as usually seen in other RTX compounds. Figure 34 displays the temperature dependence of $\Delta S_M$ for crystalline RCuAl (R = Gd-Er) compounds under a magnetic field change of 5 T. It can be seen that the $\Delta S_M$ value increases gradually with the R atom changing from Gd to Er, and reaches the maximum as high as 23.9 J/kg K for HoCuAl. The large MCE of RCuAl (R = Gd-Er) compounds suggests them as promising candidates of magnetocaloric materials in low temperature range. In addition, Dong et al.[89, 90] compared the magnetic and magnetocaloric properties of amorphous and crystalline RCuAl (R = Gd, Tb) alloys. Figure 35 shows the temperature dependence of magnetization for (a) amorphous and (b) crystalline TbCuAl alloy under 0.1 T.[90] The crystalline TbCuAl experiences a FM-PM transition at $T_C$ = 52 K while the amorphous counterpart shows a small cusp centered at 20 K, which is attributed to a spin-glass transition. Different MCE was also observed in the amorphous and crystalline TbCuAl alloys due to the different nature of magnetic transitions. As shown in Fig. 36,[90] the maximum $-\Delta S_M$ value is obtained to be 4.5 J/kg K around 36 K under a field change of 5 T for amorphous TbCuAl alloy. However, $-\Delta S_M$ peak reaches 14.4 J/kg K around 52 K under the same field change for crystalline TbCuAl alloy, which is much larger than that of amorphous alloy.

Since Dong et al.[86] did not obtain the pure HoCuAl compound, and therefore, the existence of impurity phase may influence the MCE. Later, Wang et al.[91] successfully synthesized pure HoCuAl compound and studied the magnetic properties and MCE in detail. Figure 37(a) shows the temperature dependence of magnetization for HoCuAl compound under 0.01 T.[91] The HoCuAl experiences a FM to PM transition around $T_C$ = 11.2 K, which is consistent with the data from impure sample (12 K).[86] The ZFC and FC curves show a distinct discrepancy below $T_C$ as often observed in other RTX compounds. Taking into account of the magnetic anisotropy and low $T_C$ for HoCuAl, this thermomagnetic irreversibility is likely attributed to the domain-wall-pinning effect. Figure 37(b) shows the $\Delta S_M$ as a function of temperature for HoCuAl under different magnetic field changes.[91] For a relatively low field change of 2 T, the maximum $-\Delta S_M$ value for the pure HoCuAl compound is as high as 17.5 J/kg K at $T_C$ = 11.2 K, which is 25% higher than that of impure sample (14.0 J/kg K). This large $-\Delta S_M$ is also deemed as the highest $-\Delta S_M$ value in RTAl (R = Gd-Tm) system reported so far.

In 2013, Mo et al.[16] reported a low-field induced giant MCE in TmCuAl compound. With the decrease of temperature, TmCuAl exhibits a transition from PM to FM state at $T_C$ = 2.8 K, which likely corresponds to the presence of a longitudinal spin wave magnetic structure according to the neutron diffraction studies.[92] Figure 38 displays the temperature dependence of $-\Delta S_M$ and $\Delta T_{ad}$ for TmCuAl under different magnetic field changes.[16] Large reversible MCE under low field change can be observed around the $T_C$, e.g., the $-\Delta S_M$ and $\Delta T_{ad}$ values of TmCuAl are 17.2 J/kg K and 4.6 K for a field change of 2 T, respectively. Particularly, the $-\Delta S_M$ reaches as high as 12.2 J/kg K for a low field change of 1 T, which can be applied by a permanent magnet. This giant MCE without thermal and magnetic hysteresis indicates that TmCuAl could be an attractive magnetic refrigerant around the helium liquefaction temperature.

## 3.5 RPdAl compounds

It has been found that different crystal structures can be formed in RPdAl compounds depending on the variation of heat treatment technique. RPdAl compounds crystallize in a hexagonal ZrNiAl-type structure as metastable high-temperature modification (HTM) through a high-temperature (~1050℃) annealing and rapid cooling process, while they crystallize in an orthorhombic TiNiSi-type structure as stable low-temperature modification (LTM) by a low-temperature (~750 ℃) annealing process.[93] In addition, an isostructural phase transition from a high-temperature HTM I phase to a low-temperature HTM II phase was observed in GdPdAl and TbPdAl with a metastable HTM.[94, 95]

The RPdAl (R = Gd, Tb, Dy) compounds with the hexagonal ZrNiAl-type HTM were reported to exhibit two magnetic transitions with the variation of temperature.[94-96] Shen *et al*.[97] studied the MCE of HTM-TbPdAl compound by magnetization measurements. Figure 39(a) shows the temperature dependence of ZFC and FC magnetizations under a magnetic field of 0.05 T for HTM-TbPdAl compound. It can be seen that the TbPdAl undergoes a PM-AFM transition around $T_N$ = 43 K. In addition, another anomaly is observed at $T_t$ = 22 K, which is associated with an AFM structure transition of frustrated Tb moments from purely commensurate AFM to incommensurate AFM magnetic structure.[98, 99] Moreover, the discrepancy between ZFC and FC curves below 30 K is likely related to the frustration effects of the magnetic structures. Figure 39(b) shows the temperature dependence of $\Delta S_M$ for HTM-TbPdAl under different magnetic field changes.[97] A small positive $\Delta S_M$ value can be observed below $T_N$ under relatively low magnetic field changes, which is due to the disordered magnetic sublattices antiparallel to the applied magnetic field. But the $\Delta S_M$ gradually changes to negative value with the increase of magnetic field. Such a sign change of $\Delta S_M$ indicates the occurrence of field-induced AFM-FM magnetic transition, which leads to a more ordered magnetic configuration. It is found that the $\Delta S_M$–$T$ curve shows a small peak around $T_t$, which is associated with AFM structure transition of frustrated Tb moments. In addition, a large $-\Delta S_M$ value of 11.4 J/kg K for a field change of 5 T is obtained around $T_N$, which is due to the field-induced AFM-FM transition. Moreover, the $\Delta S_M$ peak expands in a wide temperature range, and thus leading to a high RC value of 350 J/kg for a field change of 5 T.

Recently, Xu *et al.*[100, 101] systematically investigated the magnetic properties and MCE of RPdAl (R = Gd-Er) compounds with different crystal structures, and found that the magnetic properties and MCE could be affected significantly by the variation of crystal structure. Two series of RPdAl compounds were annealed at 750℃ for 50 days and at 1050~1080℃ for 10~12 days, respectively. Figure 40 shows the XRD patterns and crystal structures of these two series of RPdAl compounds at room temperature.[32, 100] The XRD measurements confirm that the low-temperature annealed samples crystallize in an orthorhombic TiNiSi-type structure as stable LTM (space group *Pnma*) while the high-temperature annealed compounds crystallize in a hexagonal ZrNiAl-type structure as metastable HTM (space group $P\bar{6}2m$). In addition, it is noted that both series of RPdAl crystallize in a purely single phase. The lattice parameters and unit cell volumes determined from the Rietveld refinement are summarized in Table 3.[32, 100] It is found that the LTM-RPdAl compounds exhibit larger unit volume than that of HTM-RPdAl.

Moreover, the lattice constants and unit cell volumes decrease linearly with the R atom sweeping from Gd to Er, which is attributed to the lanthanide contraction.

Figure 41 displays the temperature dependence of magnetization under a magnetic field of 0.01 T for LTM-RPdAl compounds.[32, 100] It is found that all LTM-RPdAl (R = Gd-Er) compounds order antiferromagnetically with the decrease of temperature, and the $T_N$ is determined to be 31, 45, 21, 10, and 10 K for R = Gd, Tb, Dy, Ho, and Er, respectively. Besides, TbPdAl and ErPdAl exhibit another AFM-AFM magnetic transition around $T_t$ = 24 K and 4 K, respectively. However, the magnetic structures and properties turn to be much different in HTM-RPdAl compounds. Figure 42 shows the temperature dependence of magnetization for HTM-RPdAl compounds.[32, 100] It can be found that all compounds, except ErPdAl exhibiting a single AFM-PM transition at $T_N$ = 5 K, show two successive magnetic transitions with the variation of temperature.

It is seen from Fig. 42(a) that the HTM-GdPdAl undergoes a FM-PM transition at $T_C$ = 49 K followed by a spin reorientation at $T_{SR}$ = 16 K. In addition, it is interesting to note that HTM-GdPdAl experiences an isostructural phase transition around 180 K from a high-temperature HTM I phase to a low-temperature HTM II phase, which thus leads to the change of lattice parameters, electrical resistance, magnetic susceptibility, and etc.[95] With the decrease in temperature, HTM-TbPdAl undergoes a PM-AFM transition around $T_N$ = 43 K followed by an AFM structure transition at $T_t$ = 22 K, which is in a good agreement with the result of Shen *et al.*[97] In HTM-TbPdAl compound with hexagonal ZrNiAl-type structure, one-third of the Tb moments (Tb2) are highly frustrated in the temperature range of $T_N$ and $T_t$, and this geometrical frustration of Tb2 spins results in the AFM structure transition from a purely commensurate AFM structure at higher temperature to a purely imcommensurate AFM magnetic structure when $T < T_t$. Besides, two-third of the non-frustrated Tb moments (Tb1 and Tb3) exhibit commensurate AFM ordering below $T_N$. HTM-DyPdAl undergoes two successive magnetic phase transitions with the decrease of temperature: a PM-FM transition at $T_C$ = 22 K followed by a spin reorientation transition at $T_{SR}$ = 14 K. These transition temperatures are a little lower than the data ($T_C$ = 25 K and $T_{SR}$ = 17 K) obtained by AC susceptibility and electrical resistance measurements.[96, 102] The isostructural phase transition from HTM I phase to HTM II phase is not observed in HTM-DyPdAl compound. The X-ray photoelectron spectroscopy (XPS) studies reveal that the localization of the Pd 4$d$ orbitals and the hybridization processes contribute to the coupling of the system in HTM II phase, and thus the disappearance of HTM II phase is likely related to the weak molecular field and absence of such coupling.[96, 103] Talik *et al*.[102] found that the single crystal HTM-DyPdAl exhibits distinct magnetic hysteresis and several metamagnetic transitions with the applied magnetic field along *a* axis, while it behaves perfect magnetic reversibility along *c* axis. This fact indicates that HTM-DyPdAl may have a canted magnetic structure with an AFM moment component along *a* axis. It can be seen from Fig. 42(d) that HTM-HoPdAl experiences a PM-AFM transition at $T_N$ = 12 K. Besides, another transition is observed around $T_t$ = 4 K, which corresponds to an AFM-AFM transition. Similar to HTM-DyPdAl, Talik *et al*.[104] found that the overlapping of 4$d$ orbitals of the neighboring Pd atoms in HTM-HoPdAl is also reduced due to the formation of narrow Pd 4$d$ band. Therefore, the coupling of the system in HTM II phase is weakened with the reduction of the localization of Pd 4$d$ orbitals and the hybridization processes, and thus isostructural phase transition from HTM I phase to HTM II phase can not occur in HTM-HoPdAl compound. In addition, the magnetic measurements on single crystal reveal that HTM-HoPdAl exhibits high magnetocrystalline anisotropy. The AFM characteristic can be found in the thermal variation of magnetization along *a* axis, and it correlates with the increase of lattice parameter *a*. Along the *c* axis, HTM-HoPdAl behaves a FM characteristic accompanied by a shrinkage of lattice parameter *c*.

Figure 43 shows the magnetization isotherms and the temperature dependence of $\Delta S_M$ under different magnetic field changes for HTM-RPdAl (R = Gd, Tb, and Dy) compounds.[32, 100] The magnetization of HTM-GdPdAl below $T_C$ increases rapidly at low fields and tends to saturate with increasing field, corresponding to the typical FM behavior. In addition to the $\Delta S_M$ peak at $T_C$, another $\Delta S_M$ peak can be observed around $T_{SR}$ = 16 K. These two successive peaks result in a table-like $\Delta S_M$–$T$ curve between $T_C$ and $T_{SR}$, which is favorable to the Ericsson-cycle magnetic refrigeration. For the field changes of 2 and 5 T, the maximum –$\Delta S_M$ values of HTM-GdPdAl are 5.3 and 9.2 J/kg K, respectively. A field-induced metamagnetic transition from AFM to FM states is observed below $T_N$ in HTM-TbPdAl as seen in ref.[97], and the negative slope of Arrott plots below $T_N$ confirms that the nature of this metamagnetic transition is of first-order. The critical field for metamagnetic transition is obtained to be 0.55 T at 9 K, and this relatively low critical field suggests the possible large MCE in HTM-TbPdAl compound. The maximum –$\Delta S_M$ values for the magnetic field changes of 2 and 5 T are 5.8 and 11.4 J/kg K, respectively, which are consistent with the results of Shen *et al*.[97] Compared with GdPdAl and HoPdAl, the HTM-DyPdAl is much harder to be magnetized to saturation even with the application of 14 T. Since the PM-FM and SR transitions take place closely, the two $\Delta S_M$ peaks overlap with each other and then lead to a single $\Delta S_M$ peak. Similar phenomenon has also been observed in other systems, such as ErGa[105] and Ho$_3$Al$_2$.[106] For the field changes of 2, 5, and 7 T, the maximum $-\Delta S_M$ values of HTM-DyPdAl are 7.8, 14.7, and 17.8 J/kg K, respectively. The $\Delta S_M$ peak broadens asymmetrically towards high temperature with increasing field, suggesting the possible presence of short-range FM correlations in the PM region.

The magnetization isotherms of LTM-RPdAl and HTM-RPdAl (R = Ho and Er) under applied fields up to 7 T are displayed in Fig. 44.[100, 101] For LTM-RPdAl (R = Ho and Er) compounds, a field-induced metamagnetic transition from AFM to FM states can be clearly seen below $T_N$ with the increase of magnetic field. On the contrary, HTM-RPdAl (R = Ho and Er) compounds exhibit a weak metamagnetic transition at lower fields and then tend to saturation with a higher magnetization in comparison with that of LTM-RPdAl. For example, the $H_{cr}$ and $M_{7\,T}$ at 2 K are 1.5 T and 141 A/m$^2$ kg for LTM-HoPdAl, and 0.15 T and 163 A/m$^2$ kg for HTM-HoPdAl, respectively.[101] This fact indicates that the AFM coupling in LTM-RPdAl is much stronger than that in HTM-RPdAl compounds. Figure 45 shows the Arrott plots of LTM-RPdAl and HTM-RPdAl (R = Ho and Er) compounds.[100, 101] According to Banerjee criterion,[107] a magnetic transition is considered as first-order when the Arrott plots exhibit negative slope or inflection point; otherwise it is expected to be of second-order when the Arrott plots exhibit positive slope. It is clearly seen that the slope of Arrott plots below $T_N$ is negative for all compounds, further confirming the first-order AFM-FM metamagnetic transition. On the other hand, the positive slope of all Arrott plots above $T_N$ proves the nature of a field-induced second-order PM-FM transition. The temperature dependence of $\Delta S_M$ for LTM-RPdAl and HTM-RPdAl (R = Ho and Er) compounds under different magnetic field changes are shown in Fig. 46.[100, 101] A small positive value of $\Delta S_M$ is observed at low temperatures and low fields, but it gradually changes to negative value with the increase of magnetic field. Such a sign change of $\Delta S_M$ is due to the field-induced metamagnetic transition from AFM to FM states. For the field changes of 2, 5, and 7 T, the maximum $-\Delta S_M$ values are 12.8, 20.6, and 23.6 J/kg K for of HTM-HoPdAl, and 12.0, 24.3, and 28.4 J/kg K for HTM-ErPdAl, respectively. These $\Delta S_M$ values are comparable with or even larger than those of some magnetocaloric materials in the same temperature range.[67, 79, 108-110] However, the maximum $-\Delta S_M$ values for a field change of 5 T are 13.7 and 11.6 J/kg K for LTM-HoPdAl and LTM-ErPdAl compounds, respectively. This lower MCE is attributed to the strong AFM coupling in LTM-RPdAl compounds. Consequently, the MCE of LTM-RPdAl and HTM-RPdAl are summarized in Table 2. The relatively high $\Delta S_M$ and RC values for HTM-RPdAl suggest them as promising materials for magnetic refrigeration in low temperature range.

## 4. Conclusions

The MCE of RTX (R = Gd－Tm, T = Fe－Cu and Pd, X = Al and Si) intermetallic compounds have been reviewed, and it is found that these compounds exhibit various magnetic properties and different MCE in a relatively low temperature range. Generally, the crystal structure and the magnetic ground state will not change with the variation of R atom from Gd to Tm. However, the transition temperature decreases with the R atom sweeping from Gd to Tm, indicating that the magnetic interactions (mainly RKKY interaction) may be strengthened. Meanwhile, the MCE increases with varying R atom from Gd to Tm and usually reaches the maximum value in Ho-based compounds, which is likely related to the high angular momentum $J$ of $Ho^{3+}$ ion. Usually, the T and X atoms do not contribute the magnetic moments due to the hybridization between $d$ states of T and $p$ states of X atoms. However, the change of either T atom or X atom would result in the different crystal and magnetic structures, and then affect the MCE of RTX compounds. In addition, the crystal structures and magnetic properties of some RTX, such as RCuSi, RFeAl, and RPdAl, can be also influenced by different treatment processes, and then leading to a large difference in MCE. Particularly, it is worthwhile to note that some of RTX compounds exhibit large reversible MCE especially under low magnetic field change, making them promising candidates for magnetic refrigeration materials in low temperature range, such as the liquefaction temperatures of helium, hydrogen, and nitrogen.

Although the MCE of many RTX compounds have been investigated systematically in the past decades, a lot of important problems, such as the regulatory mechanism of giant MCE, the relationship between heat treatment and physical properties, and etc, have not been addressed well. In addition, other key issues which are closely related to the practical application of magnetic refrigeration, such as mechanical property, corrosion behavior, and preparation technique, and etc, are also need to be taken into account. Consequently, previous studies have revealed that RTX compounds exhibit large MCE in low temperature range, suggesting them as promising materials for low-temperature magnetic refrigeration. However, we still have a long way to go to fulfill the practical application of these materials in magnetic refrigeration.

**References:**

1.  Tishin A M and Spichkin Y I 2003 *in The Magnetocaloric Effect and its Applications*, edited by Coey J M D, Tilley D R and Vij D R (Bristol: IOP Publishing).

2.  Gschneidner K A Jr, Pecharsky V K and Tsokol A O 2005 *Rep. Prog. Phys.* **68** 1479.

3.  Shen B G, Sun J R, Hu F X, Zhang H W and Cheng Z H 2009 *Adv. Mater.* **21** 4545.

4.  Pecharsky V K and Gschneidner K A Jr 1997 *Phys. Rev. Lett.* **78** 4494.

5.  Hu F X, Shen B G, Sun J R and Zhang X X 2000 *Chin. Phys.* **9** 550.

6.  Hu F X, Shen B G, Sun J R, Cheng Z H, Rao G H and Zhang X X 2001 *Appl. Phys. Lett.* **78** 3675.

7.  Wada H and Tanabe Y 2001 *Appl. Phys. Lett.* **79** 3302.

8.  Guo Z B, Du Y W, Zhu J S, Huang H, Ding W P and Feng D 1997 *Phys. Rev. Lett.* **78** 1142.

9.  Tegus O, Brück E, Buschow K H J and de Boer F R 2002 *Nature* **415** 150.

10. Liu J, Gottschall T, Skokov K P, Moore J D and Gutfleisch O 2012 *Nat. Mater.* **11** 620.

11. Shen B G, Hu F X, Dong Q Y and Sun J R 2013 *Chin. Phys. B* **22** 017502.

12. Hu F X, Shen B G and Sun J R 2013 *Chin. Phys. B* **22** 037505.

13. Zhang H, Hu F X, Sun J R and Shen B G 2013 *Sci. China-Phys. Mech. Astron.* **56** 2302.

14. Zhang H, Shen B G, Xu Z Y, Shen J, Hu F X, Sun J R and Long Y 2013 *Appl. Phys. Lett.* **102** 092401.

15. Chen J, Shen B G, Dong Q Y, Hu F X and Sun J R 2010 *Appl. Phys. Lett.* **96** 152501.

16. Mo Z J, Shen J, Yan L Q, Wu J F, Wang L C, Lin J, Tang C C and Shen B G 2013 *Appl. Phys. Lett.* **102** 192407.

17. Gupta S and Suresh K G 2015 *J. Alloys Compd.* **618** 562.

18. Bodak O I, Gladyshevskii E I and Kripyakevich P I 1970 *Zh. Strukt. Khim.* **11** 283.

19. Welter R, Venturini G and Malaman B 1992 *J. Alloys Compd.* **189** 49.

20. Napoletano M, Canepa F, Manfrinetti P and Merlo F 2000 *J. Mater. Chem.* **10** 1663.

21. Zhang H, Sun Y J, Niu E, Yang L H, Shen J, Hu F X, Sun J R and Shen B G 2013 *Appl. Phys. Lett.* **103** 202412.

22. Zhang H, Sun Y J, Yang L H, Niu E, Wang H S, Hu F X, Sun J R and Shen B G 2014 *J. Appl. Phys.* **115**, 063901.

23. Gschneidner K A Jr, Pecharsky V K, Pecharsky A O and Zimm C B 1999 *Mater. Sci. Forum* **315-317** 69.

24. Wang J L, Marquina C, Ibarra M R and Wu G H 2006 *Phys. Rev. B* **73** 094436.

25. Ouyang Z W, Pecharsky V K, Gschneidner K A Jr, Schlagel D L and Lograsso T A 2006 *Phys. Rev. B* **74** 094404.

26. Pecharsky V K and Gschneidner K A Jr 1999 *J. Appl. Phys.* **86** 565.

27. Smaili A and Chahine R 1997 *J. Appl. Phys.* **81** 824.

28. Zhang X X, Zhang B, Yu S, Liu Z H, Xu W J, Liu G D, Chen J L, Cao Z X and Wu G H 2007 *Phys. Rev. B* **76** 132403.

29. Welter R, Venturini G, Ressouche E and Malaman B 1994 *J. Alloys Compd.* **210** 279.

30. Ijjaali I, Welter R, Venturini G and Malaman B 1999 *J. Alloys Compd.* **292** 4.

31. Leciejewicz J, Stuesser N, Kolenda M, Szytula A and Zygmunt A 1996 *J. Alloys Compd.* **240** 164.

32. Xu Z Y 2012 *Mangetism and magnetocaloric effects in rare earth-transition metal compounds with low temperature transition* (Ph.D. thesis) (Graduate University of Chinese Academy of Sciences) (in Chinese).

33. Gupta S and Suresh K G 2013 *Mater. Lett.* **113** 195.

34. Bodak O I, Yarovetz V I and Gladyshevskij E I 1974 *in Tesizy Dokl. Tvet. Vses. Konf. kristallokhim. Intermet. Soedin, 2nd ed.*, edited by Rykhal R M (Lvov. os. Univ., Lvov, USSR, p. 32).

35. Szytula A, Balanda M, Hofmann M, Leciejewicz J, Kolenda M, Penc B and Zygmunt A 1999 *J. Magn. Magn. Mater.* **191** 122.

36. Zhang H, Wu Y Y, Long Y, Wang H S, Zhong K X, Hu F X, Sun J R and Shen B G 2014 *J. Appl. Phys.* **116** 213902.

37. Zhang H, Li Y W, Liu E K, Ke Y J, Jin J L, Long Y and Shen B G 2015 *Sci. Rep.* **5** 11929.

38. Gupta S, Rawat R and Suresh K G 2014 *Appl. Phys. Lett.* **105** 012403.

39. Midya A, Das S N, Mandal P, Pandya S and Ganesan V 2011 *Phys. Rev. B* **84** 235127.

40. Midya A, Khan N, Bhoi D and Mandal P 2012 *Appl. Phys. Lett.* **101** 132415.

41. Zou J D, Shen B G, Gao B, Shen J and Sun J R 2009 *Adv. Mater.* **21** 693.

42. Franco V, Blázquez J S and Conde A 2006 *Appl. Phys. Lett.* **89** 222512.

43. Franco V, Blázquez J S, Ingale B and Conde A 2012 *Annu. Rev. Mater. Res.* **42** 305.

44. Singh N K, Pecharsky V K and Gschneidner K A Jr 2008 *Phys. Rev. B* **77** 054414.

45. Rieger W and Parthé E 1969 *Monatsch. Chem.* **100** 439.

46. Iandelli A 1983 *J. Less-Common Met.* **90** 121.

47. Mugnoli A, Albinati A and Hewat A W 1984 *J. Less-Common Met.* **97** L1.

48. Kido H, Hoshikawa T, Shimada M and Koizumi M 1983 *Phys. Status Solidi A* **77** K121.

49. Bażela W, Szytuła A and Leciejewicz J 1985 *Solid State Commun.* **56** 1043.

50. Oleś A, Duraj R, Kolenda M, Penc B and Szytula A 2004 *J. Alloys Compd.* **363** 63.

51. Schobinger-Papamantellos P, Buschow K H J and Ritter C 2004 *J. Alloys Compd.* **384** 12.

52. Schobinger-Papamantellos P, Buschow K H J, Duong N P and Ritter C 2001 *J. Magn. Magn. Mater.* **223** 203.

53. Chen J, Shen B G, Dong Q Y and Sun J R 2010 *Solid State Commun.* **150** 1429.

54. Chen J 2010 *Magnetic properties and magnetocaloric effects in $Ni_2$In-type RCuSi and CrB-type RGa compounds* (Ph.D. thesis) (Graduate University of Chinese Academy of Sciences) (in Chinese).

55. Samanta T, Das I and Banerjee S 2007 *Appl. Phys. Lett.* **91** 152506.

56. Wang H T, Zhou X W, Sun L B, Dong J L and Yu S Y 2009 *Nucl. Eng. Des.* **239** 484.

57. Kilmczak M, Talik E, Jarosz J and Mydlarz T 2008 *Mater. Sci-Poland* **26** 953.

58. Dong Q Y, Shen B G, Chen J, Shen J, Zhang H W and Sun J R 2009 *J. Appl. Phys.* **105** 07A305.

59. Kastil J, Javorský P, Kamarád J, Diop L V B, Isnard O and Arnold Z 2014 *Intermetallics* **54** 15.

60. Li L W, Huo D X, Qian Z H and Nishimura K 2014 *Intermetallics* **46** 231.

61. Oesterreicher H 1973 *J. Phys. Chem. Solids* **34** 1267.

62. Oesterreicher H 1971 *Phys. Status Solidi A* **7** K55.

63. Mulders A M 1998 *in Complex Magnetic Phenomena in Rare Earth Intermetallic Compounds* (Ph.D. thesis) (Delft University Press, Delft).

64. Zhang Y K, Wilde G, Li X, Ren Z M and Li L W 2015 *Intermetallics* **65** 61.

65. Oesterreicher H 1971 *J. Less-Common Met.* **25** 228.

66. Jarosz J, Talik E, Mydlarz T, Kusz J, Böhm H and Winiarski A 2000 *J. Magn. Magn. Mater.* **208** 169.

67. Zhang X X, Wang F W and Wen G H 2001 *J. Phys.: Condens. Matter* **13** L747.

68. Gschneidner K A Jr and Pecharsky V K 1997 *in Rare Earth: Science, Technology and Applications III*, edited by R. G. Bautista, C. O. Bounds, T. W. Ellis, and B. T. Kilbourn (PA: The Mineral, Metals and Materials Society, Warrendale).

69. Chelvane J A, Das T, Mahato R N, Morozkin A V, Lamsal J, Yelon W B, Nirmala R and Malik S K 2010 *J. Appl. Phys.* **107** 09A906.

70. Mo Z J, Shen J, Yan L Q, Tang C C, Wang L C, Wu J F, Sun J R and Shen B G 2015 *Intermetallics* **56** 75.

71. Javorský P, Burlet P, Ressouche E, Sechovský V and Lapertot G 1996 *J. Magn. Magn. Mater.* **159** 324.

72. Korte B J, Pecharsky V K and Gschneidner K A Jr 1998 *J. Appl. Phys.* **84** 5677.

73. Danis S, Javorský P and Rafaja D 2002 *J. Alloys Compd.* **345** 10.

74. Si L, Ding J, Li Y, Yao B and Tan H 2002 *Appl. Phys. A* **75** 535.

75. Ehlers G and Maletta H 1997 *Z. Phys. B* **101** 317.

76. Javorský P, Burlet P, Sechovský V, Andreev A V, Brown J and Svoboda P 1997 *J. Magn. Magn. Mater.* **166** 133.

77. Singh N K, Suresh K G, Nirmala R, Nigam A K and Malik S K 2006 *J. Magn. Magn. Mater.* **302** 302.

78. Singh N K, Suresh K G, Nirmala R, Nigam A K and Malik S K 2006 *J. Appl. Phys.* **99** 08k904.

79. Singh N K, Suresh K G, Nirmala R, Nigam A K and Malik S K 2007 *J. Appl. Phys.* **101** 093904.

80. Mo Z J 2014 The research of magnetocaloric effec in the low and medium temperature (Ph.D. thesis) (Hebei University of Technology) (in Chinese).

81. Mo Z J, Shen J, Chen G F, Yan L Q, Zheng X Q, Wu J F, Tang C C, Sun J R and Shen B G 2014 *J. Appl. Phys.* **115** 17A909.

82. Cui L, Wang L C, Dong Q Y, Liu F H, Mo Z J, Zhang Y, Niu E, Xu Z Y, Hu F X, Sun J R and Shen B G 2015 *J. Alloys Compd.* **622** 24.

83. Wang L C, Dong Q Y, Lu J, Shao X P, Mo Z J, Xu Z Y, Sun J R, Hu F X and Shen B G 2013 *J. Appl. Phys.* **114** 213907.

84. Wang L C, Dong Q Y, Xu Z Y, Hu F X, Sun J R and Shen B G 2013 *J. Appl. Phys.* **113** 023916.

85. Dong Q Y, Shen B G, Chen J, Shen J and Sun J R 2009 *J. Appl. Phys.* **105** 113902.

86. Dong Q Y, Chen J, Shen J, Sun J R and Shen B G 2012 *J. Magn. Magn. Mater.* **324** 2676.

87. Javorský P, Havela L, Sechovský V, Michor H and Jurek K 1998 *J. Alloys Compd.* **264** 38.

88. Andreev A V, Javorský P and Lindbaum A 1999 *J. Alloys Compd.* **290** 10.

89. Dong Q Y, Shen B G, Chen J, Shen J, Wang F, Zhang H W and Sun J R 2009 *Solid State Commun.* **149** 417.

90. Dong Q Y, Shen B G, Chen J, Shen J and Sun J R 2011 *Solid State Commun.* **151** 112.

91. Wang L C, Dong Q Y, Mo Z J, Xu Z Y, Hu F X, Sun J R and Shen B G 2013 *J. Appl. Phys.* **114** 163915.

92. Javorský P, Gubbens P C M, Mulders A M, Prokes K, Stüsser N, Gortenmulder T J and Hendrikx R W A 2002 *J. Magn. Magn. Mater.* **251** 123.

93. Hulliger F 1995 *J. Alloys Compd.* **218** 44.

94. Dönni A, Kitazawa H, Fischer P and Fauth F 1999 *J. Alloys Compd.* **289** 11.

95. Talik E, Skutecka M, Kusz J, Böhm H, Jarosz J, Mydlarz T and Winiarski A 2001 *J. Alloys Compd.* **325** 42.

96. Talik E, Skutecka M, Kusz J and Böhm H 2004 *J. Magn. Magn. Mater.* **272-276** 767.

97. Shen J, Xu Z Y, Zhang H, Zheng X Q, Wu J F, Hu F X, Sun J R and Shen B G 2011 *J. Magn. Magn. Mater.* **323** 2949.

98. Javorský P, Prokleška J, Isnard O and Prchal J 2008 *J. Phys.: Condens. Matter* **20** 104223.

99. Dönni A, Kitazawa H, Keller L, Fischer P, Javorsky P, Fauth F and Zolliker M 2009 *J. Alloys Compd.* **477** 16.

100. Xu Z Y, Shen J and Shen B G 2013 *Rare Mater. Lett.* **32** 84 (in Chinese).

101. Xu Z Y and Shen B G 2012 *Sci. China Technol. Sci.* **55** 445.

102. Talik E, Skutecka M, Mydlarz T, Kusz J and Böhm H 2005 *J. Alloys Compd.* **391** 1.

103. Kusz J, Böhm H, Talik E, Skutecka M and Deniszczyk J 2003 *J. Alloys Compd.* **348** 65.

104. Talik E, Skutecka M, Kusz J, Bohm H and Mydlarz T 2003 *J. Alloys Compd.* **359** 103.

105. Chen J, Shen B G, Dong Q Y, Hu F X and Sun J R 2009 *Appl. Phys. Lett.* **95** 132504.

106. Zhang H, Xu Z Y, Zheng X Q, Shen J, Hu F X, Sun J R and Shen B G 2012 *Solid State Commun.* **152** 1127.

107. Banerjee S K 1964 *Phys. Lett.* **12** 16.

108. Shen J, Zhao J L, Hu F X, Wu J F, Sun J R and Shen B G 2010 *Chin. Phys. B* **19** 047502.

109. Hu W J, Du J, Li B, Zhang Q and Zhang Z D 2008 *Appl. Phys. Lett.* **92** 192505.

110. von Ranke P J, Mota M A, Grangeia D F, Carvalho A M G, Gandra F C G, Coelho A A, Caldas A, de Oliveira N A and Gama S 2004 *Phys. Rev. B* **70** 134428.

Table 1, Magnetocaloric properties of RTSi (R = Gd－Er, T = Fe－Cu) compounds.

| Materials | Ground State | $T_{ord}$ (K) | $-\Delta S_M$ (J/kg K) | | $\Delta T_{ad}$ (K) | | RC (J/kg) | Refs. |
|---|---|---|---|---|---|---|---|---|
| | | | 2 T | 5 T | 2 T | 5 T | 5 T | |
| GdFeSi | FM | 130 | 6.0 | 11.3 | - | - | 373 | This work |
| TbFeSi | FM | 110 | 9.8 | 17.5 | 4.1 | 8.2 | 311 | [21] |
| DyFeSi | FM | 70 | 9.2 | 17.4 | 3.4 | 7.1 | 308 | [21] |
| HoFeSi | FM+ AFM/FIM | 29, 20 | 7.1, -5.6 | 16.2, -6.0 | - | - | 309, -50 | [22] |
| ErFeSi | FM | 22 | 14.2 | 23.1 | 2.9 | 5.7 | 365 | [14] |
| HoCoSi | FM | 14 | 13 | 20.5 | 3.1 | - | 410 | [33] |
| ErCoSi | FM | 5.5 | 18.7 | 25.0 | - | - | 372 | [32] |
| TbNiSi | AFM | 14.8 | 1.7 | 9.4 | - | - | 182 | This work |
| DyNiSi | AFM | 8.8 | 12.1 | 22.9 | - | - | 434 | [37] |
| HoNiSi | AFM | 3.8 | 17.5 | 26.0 | 4.5 | 8.5 | 471 | [36] |
| ErNiSi | AFM | 3.2 | 8.8 | 19.0 | 2.5[38] | - | 309 | This work |
| GdCuSi | AFM | 14 | 2.6 | 9.2 | - | - | 194[a] | [54] |
| TbCuSi | AFM | 11 | 2.7 | 10.0 | - | - | 246[a] | [54] |
| DyCuSi | AFM | 10 | 10.5 | 24.0 | - | - | 381 | [53] |
| HoCuSi | AFM | 7 | 16.7 | 33.1 | - | - | 385 | [15] |
| ErCuSi | AFM | 7 | 14.5 | 23.1 | - | - | 471[a] | [54] |

[a]The RC values were estimated from the temperature dependence of $\Delta S_M$ in the reference literatures.

Table 2, Magnetocaloric properties of RTAl (R = Gd－Tm, T = Fe－Cu and Pd) compounds.

| Materials | Ground State | $T_{ord}$ (K) | $-\Delta S_M$ (J/kg K) | | $\Delta T_{ad}$ (K) | | RC (J/kg) | Refs. |
|---|---|---|---|---|---|---|---|---|
| | | | 2 T | 5 T | 2 T | 5 T | 5 T | |
| GdFeAl | FIM | 265 | 1.8 | 3.7 | - | - | 420 | [58] |
| TbFeAl | FIM | 196 | - | 3.3[a] | 0.8 | 1.6[a] | 268[a] | [59] |
| DyFeAl | FM | 129 | 3.1 | 6.4 | - | - | 457[b] | [60] |
| HoFeAl | FM | 80 | 3.4 | 7.5 | - | - | 435 | [64] |
| ErFeAl | FM | 55 | 2.4 | 6.1 | - | - | 240 | [64] |
| GdCoAl | FM | 100 | 4.9 | 10.4 | - | - | 590[b] | [67] |
| TbCoAl | FM | 70 | 5.3 | 10.5 | - | - | 407[b] | [67] |
| DyCoAl | FM | 37 | 9.2 | 16.3 | - | - | 487[b] | [67] |
| HoCoAl | FM | 10 | 12.5 | 21.5 | - | - | 454[b] | [67] |
| TmCoAl | FM | 7.5 | 10.2 | 18.2 | - | - | 211 | [70] |
| GdNiAl | FM+AFM | 68, 30.4+15 | 5.4 | 10.9 | - | 4.1[72] | 534[b] | [74] |
| TbNiAl | FM+AFM | 48, 23 | 7.1 | 13.8 | - | - | 494 | [77] |
| DyNiAl | FM+AFM | 30, 15 | 10.0 | 19.0 | 3.5[c] | 7.0[c] | 492[b] | [78] |
| HoNiAl | FM+AFM | 14, 5 | 12.3 | 23.6 | 4 | 8.7 | 421[b] | [79] |
| ErNiAl | AFM | 6 | - | 21.6 | - | 6.3 | 230[b] | [72] |
| TmNiAl | AFM | 4 | 5.5 | 12.7 | - | - | 109[b] | [80, 81] |
| GdCuAl | FM | 81 | 5.2 | 10.1 | - | - | 460 | [89] |
| TbCuAl | FM | 52 | 6.2 | 14.4 | - | - | 401[b] | [90] |
| DyCuAl | FM | 28 | 10.9 | 20.4 | 3.6 | 7.7 | 423 | [85] |
| HoCuAl | FM | 11.2 | 17.5 | 30.6 | - | - | 486 | [91] |
| ErCuAl | FM | 7 | 14.7 | 22.9 | - | - | 321 | [86] |
| TmCuAl | FM | 2.8 | 17.2 | 24.3 | 4.6[c] | 9.4[c] | 372 | [16] |
| HTM-GdPdAl | FM | 49 | 5.3 | 6.2 | - | - | 362 | [100] |
| HTM-TbPdAl | AFM | 43 | 5.8 | 11.4 | - | - | 350 | [97] |
| HTM-DyPdAl | FM | 22 | 7.8 | 14.7 | - | - | 304 | [100] |
| LTM-HoPdAl | AFM | 10 | 2.6 | 13.7 | - | - | 174[b] | [100, 101] |
| HTM-HoPdAl | AFM | 12 | 12.8 | 20.6 | - | - | 386 | [100, 101] |
| LTM-ErPdAl | AFM | 10 | 2.0 | 11.6 | - | - | 139[b] | [100] |
| HTM-ErPdAl | AFM | 5 | 12.0 | 24.3 | - | - | 299 | [100] |

[a] $\mu_0 H = 4$ T

[b] The $RC$ values were estimated from the temperature dependence of $\Delta S_M$ in the reference literatures.

[c] The $\Delta T_{ad}$ values were calculated by using the equation $\Delta T_{ad} = -\Delta S(T, H) \times T / C_P(T, H_0)$, where $C_P(T, H_0)$ is zero-field heat capacity.

Table 3, The lattice parameters and unit cell volumes of LTM- and HTM-RPdAl compounds determined from the Rietveld refinement.[32, 100]

| RPdAl | | $a$/nm | $b$/nm | $c$/nm | $V$/nm$^3$ |
|---|---|---|---|---|---|
| R = Gd | LTM | 0.69668(0) | 0.44477(9) | 0.77586(1) | 0.2404(1) |
|        | HTM | 0.72002(2) |            | 0.40265(1) | 0.1807(8) |
| R = Tb | LTM | 0.69152(7) | 0.44274(6) | 0.77422(5) | 0.2370(4) |
|        | HTM | 0.71816(8) |            | 0.39973(9) | 0.1785(5) |
| R = Dy | LTM | 0.68823(5) | 0.44153(4) | 0.77332(8) | 0.2349(9) |
|        | HTM | 0.71893(4) |            | 0.39593(1) | 0.1772(2) |
| R = Ho | LTM | 0.68527(4) | 0.44037(9) | 0.77242(8) | 0.2331(0) |
|        | HTM | 0.71857(5) |            | 0.39320(9) | 0.1758(3) |
| R = Er | LTM | 0.68142(7) | 0.43868(5) | 0.77087(9) | 0.2304(4) |
|        | HTM | 0.71830(0) |            | 0.39080(4) | 0.1746(2) |

**Figure captions**

**Figure 1.** Temperature dependence of ZFC and FC magnetizations under 0.05 T for $R$FeSi (R = Gd-Er) compounds. The inset shows a close view of the $M$-$T$ curves of TbFeSi and DyFeSi in the PM state.[14, 21, 22]

**Figure 2.** (a) Temperature dependence of $\Delta S_M$ for $R$FeSi ($R$ = Tb and Dy) under different magnetic field changes up to 5 T. (b) Temperature dependence of calculated $\Delta S_M$ for (Tb$_{1-x}$Dy$_x$)FeSi ($x$ = 0-1) compounds and the composite material under a magnetic field change of 1 T.[21]

**Figure 3.** Temperature dependence of (a) $\Delta S_M$ and (b) $\Delta T_{ad}$ for ErFeSi under different magnetic field changes.[14]

**Figure 4.** (a) Magnetization isotherms of HoFeSi compound in the temperature range of 8-24 K. The inset shows the fraction of FM phase as a function of temperature estimated from magnetization isotherms. (b) Temperature dependence of magnetic entropy change $\Delta S_M$ for HoFeSi compound under different magnetic field changes up to 5 T.[22]

**Figure 5.** Heat capacity ($C_P$) curves for HoCoSi under different magnetic fields. The inset shows the temperature dependence of ZFC and FC curves for HoCoSi under the magnetic field of 0.01 T.[32]

**Figure 6.** Temperature dependence of (a) $\Delta S_M$ and (b) $\Delta T_{ad}$ for HoCoSi under different magnetic field changes.[32]

**Figure 7.** Temperature dependence of ZFC and FC magnetizations under 0.01 T for (Ho$_{1-x}$Er$_x$)CoSi compounds. The inset shows the transition temperatures as a function of as a function of $x$.[32]

**Figure 8.** Temperature dependence of $\Delta S_M$ for $(Ho_{1-x}Er_x)CoSi$ compounds under a magnetic field change of 2 T.[32]

**Figure 9.** Temperature dependence of $\Delta S_M$ for $Ho_{0.8}Dy_{0.2}CoSi$ compound under different magnetic field changes.[32]

**Figure 10.** Temperature dependence of ZFC and FC magnetizations for DyNiSi at 0.05 T along the parallel and perpendicular directions, respectively. The inset shows the magnetization as a function of rotation angle at 20 K under 0.05 T.[37]

**Figure 11.** Temperature dependence of $\Delta S$ for DyNiSi for different magnetic field changes along (a) parallel and (b) perpendicular directions, respectively. (c) The difference of $\Delta S$ for DyNiSi between different directions as a function of temperature for different magnetic field changes.[37]

**Figure 12.** The $\Delta S^R(\theta)$ of DyNiSi as a function of rotation angle for different magnetic field changes. The inset describes the rotation of sample from perpendicular (90°) to parallel (0°) direction in magnetic field.[37]

**Figure 13.** Temperature dependence of ZFC and FC magnetizations for (a) HoNiSi[36] and (b) ErNiSi under 0.05 T. The inset shows the temperature dependence of magnetization in various magnetic fields for HoNiSi[36] and ErNiSi, respectively.

**Figure 14.** Temperature dependence of $\Delta S_M$ for (a) HoNiSi[36] and (b) ErNiSi under different magnetic field changes up to 5 T. The $\Delta S_M$ of HoNiSi was calculated from magnetizations (open symbols) and heat capacity measurements (full symbols). The inset shows the universal curve of $\Delta S_M$ for HoNiSi compound under various magnetic field changes.

**Figure 15.** Temperature dependence of $\Delta T_{ad}$ for HoNiSi under different magnetic field changes.[36]

**Figure 16.** Temperature dependence of ZFC and FC magnetizations under the magnetic field of 0.01 T for RCuSi (R = Gd, Tb, Dy, and Er) compounds.[54]

**Figure 17.** Temperature dependence of $\Delta S_M$ for RCuSi (R = Gd, Tb, Dy, and Er) compounds under different magnetic field changes.[54]

**Figure 18.** Temperature dependence of magnetization for HoCuSi under various magnetic fields.[15]

**Figure 19.** The $\Delta S_M$ of HoCuSi as a function of temperature for different magnetic field changes.[15]

**Figure 20.** Temperature dependence of thermal expansion data ($\Delta L/L_{(50\ K)}$) under different fields for HoCuSi.[54]

**Figure 21.** (a) Temperature dependence of magnetization for GdFeAl under a field of 0.1 T. (b) The magnetization curve at 5 K for GdFeAl compound.[58]

**Figure 22.** Temperature dependence of $\Delta S_M$ for GdFeAl under the magnetic field changes of 2 and 5 T, respectively.[58]

**Figure 23.** The $T_C$ and $\Delta S_M$ for a field change of 5 T as a function of R atom for RCoAl compounds.

**Figure 24.** (a) Temperature dependence of ZFC and FC magnetizations for TmCoAl compound under the magnetic field of 0.01 T. (b) Temperature variation of the inverse dc susceptibility fitted to the Curie-Weiss law for TmCoAl.[70]

**Figure 25.** Temperature dependence of $\Delta S_M$ for TmCoAl under different magnetic field changes.[70]

**Figure 26.** Isothermal magnetization curves of TmNi$_{1-x}$Cu$_x$Al compounds as a function of magnetic field measured at 2 K in applied fields up to 5 T.[81]

**Figure 27.** The $\Delta S_M$ as a function of temperature for TmNi$_{1-x}$Cu$_x$Al compounds under a magnetic field change of 2 T.[81]

**Figure 28.** Temperature dependence of magnetization for HoNi$_{1-x}$Cu$_x$Al compounds under the magnetic field of 0.01 T.[82]

**Figure 29.** Temperature dependence of $\Delta S_M$ for HoNi$_{1-x}$Cu$_x$Al with $x =$ (a) 0.3 and (b) 0.8 under different magnetic field changes.[82]

**Figure 30.** Temperature dependence of magnetization for ErNi$_{1-x}$Cu$_x$Al compounds with $x = 0.2$, 0.5, and 0.8, respectively.[83]

**Figure 31.** Temperature dependences of real part of ac magnetic susceptibility for (a) $x = 0.5$ and (b) $x = 0.8$ samples, respectively. The inset shows the corresponding enlarged part of peak position at different frequencies for $x = 0.8$ sample.[83]

**Figure 32.** Temperature dependence of $\Delta S_M$ under different magnetic field changes for ErNi$_{1-x}$Cu$_x$Al with $x = 0.2$, 0.5, and 0.8, respectively.[83]

**Figure 33.** Temperature dependence of ZFC and FC magnetizations for crystalline RCuAl (R = Gd-Er) compounds under 0.1 and 0.05 T.[85, 86, 89, 90]

**Figure 34.** Temperature dependence of $\Delta S_M$ for crystalline RCuAl (R = Gd-Er) compounds under a magnetic field change of 5 T. [85, 86, 89, 90]

**Figure 35.** The temperature dependence of magnetization for (a) amorphous and (b) crystalline TbCuAl alloys under 0.1 T.[90]

**Figure 36.** Temperature dependence of $\Delta S_M$ for amorphous and crystalline TbCuAl alloys under different magnetic field changes.[90]

**Figure 37.** (a) Temperature dependence of magnetization for HoCuAl compound under 0.01 T. (b) The $\Delta S_M$ as a function of temperature for HoCuAl under different magnetic field changes.[91]

**Figure 38.** Temperature dependence of (a) $\Delta S_M$ and (b) $\Delta T_{ad}$ for TmCuAl under different magnetic field changes.[16]

**Figure 39.** (a) Temperature dependence of ZFC and FC magnetizations under a magnetic field of 0.05 T for HTM-TbPdAl compound. (b) Temperature dependence of $\Delta S_M$ for HTM-TbPdAl under different magnetic field changes.[97]

**Figure 40.** XRD patterns and crystal structures of (a) LTM- RPdAl and (b) HTM-RPdAl compounds at room temperature, respectively.[32, 100]

**Figure 41.** Temperature dependence of magnetization under a magnetic field of 0.01 T for LTM-RPdAl compounds.[32, 100]

**Figure 42.** Temperature dependence of magnetization under low magnetic field for HTM-RPdAl compounds.[32, 100]

**Figure 43.** Magnetization isotherms and the temperature dependence of $\Delta S_M$ under different magnetic field changes for HTM-RPdAl (R = Gd, Tb, and Dy) compounds.[32, 100]

**Figure 44.** Magnetization isotherms of LTM-RPdAl and HTM-RPdAl (R = Ho and Er) under applied fields up to 7 T.[100, 101]

**Figure 45.** Arrott plots of LTM-RPdAl and HTM-RPdAl (R = Ho and Er) compounds.[100, 101]

**Figure 46.** Temperature dependence of $\Delta S_M$ for LTM-RPdAl and HTM-RPdAl (R = Ho and Er) compounds under different magnetic field changes.[100, 101]

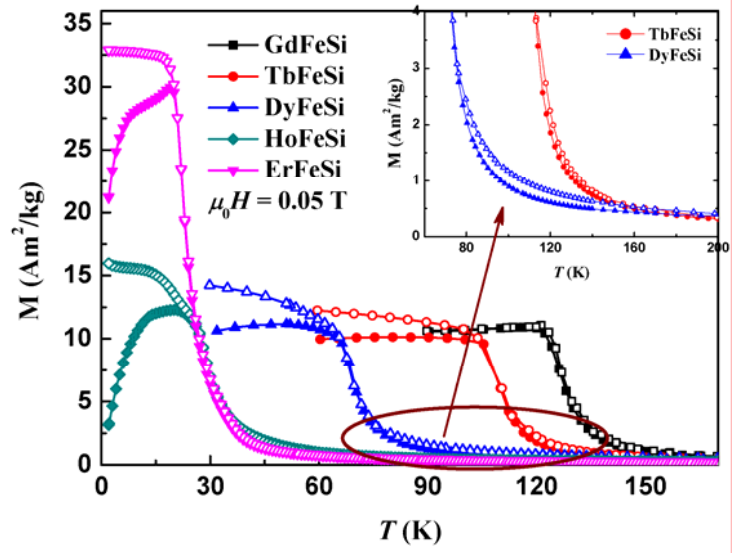

Fig. 1

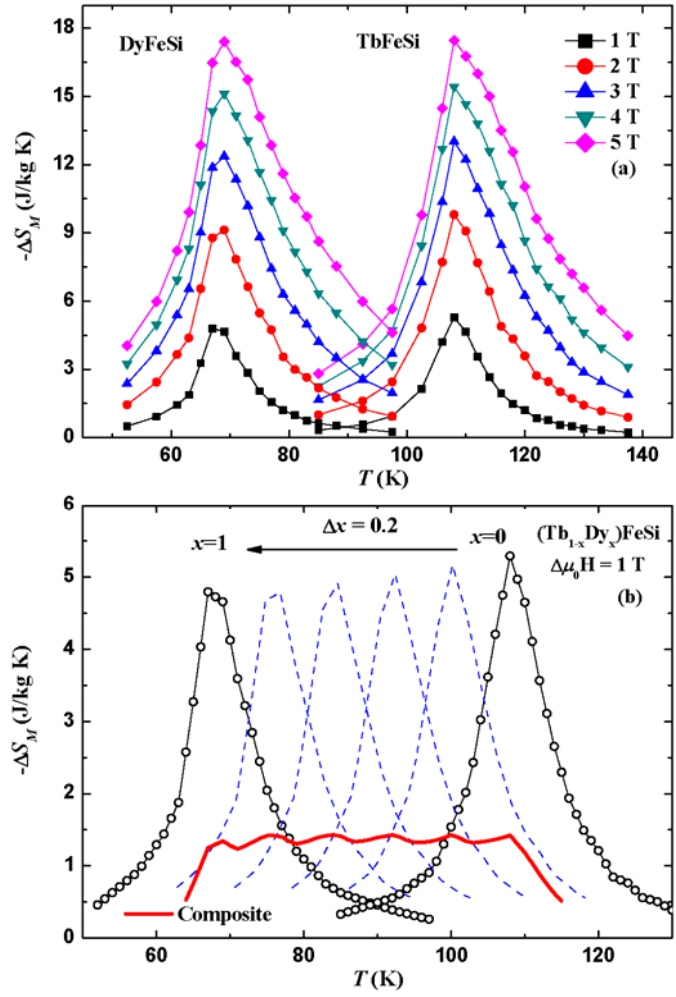

Fig. 2

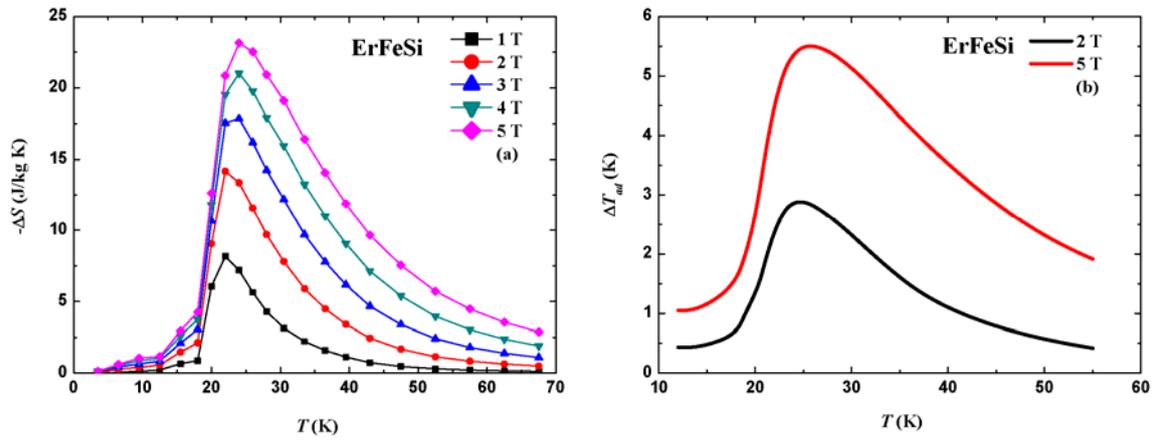

Fig. 3

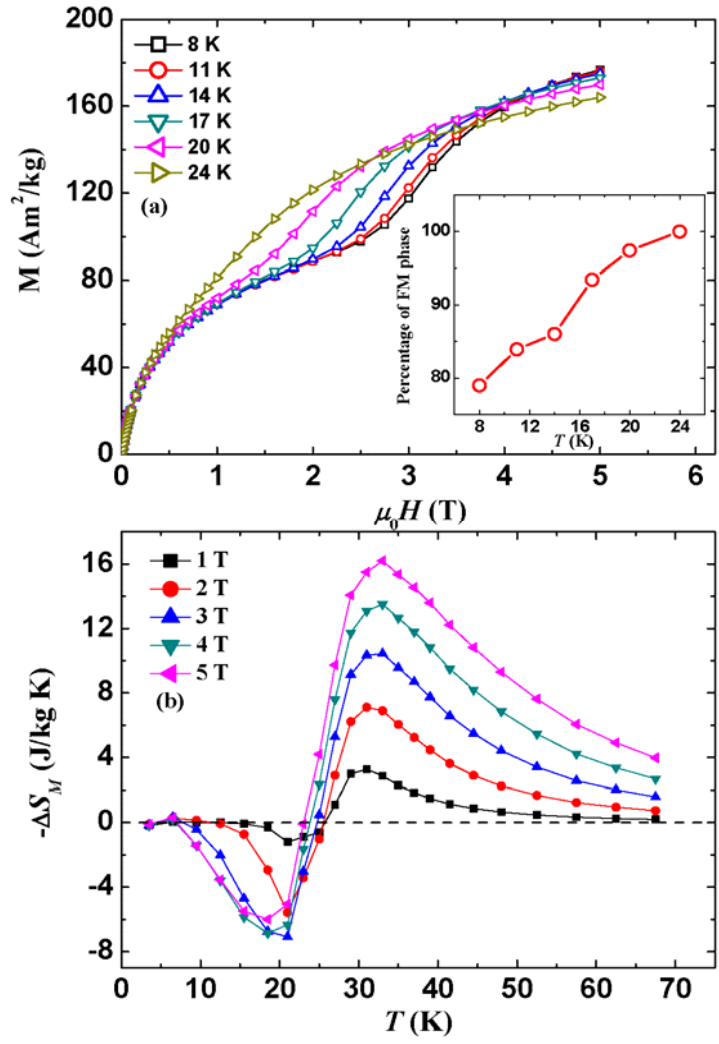

Fig. 4

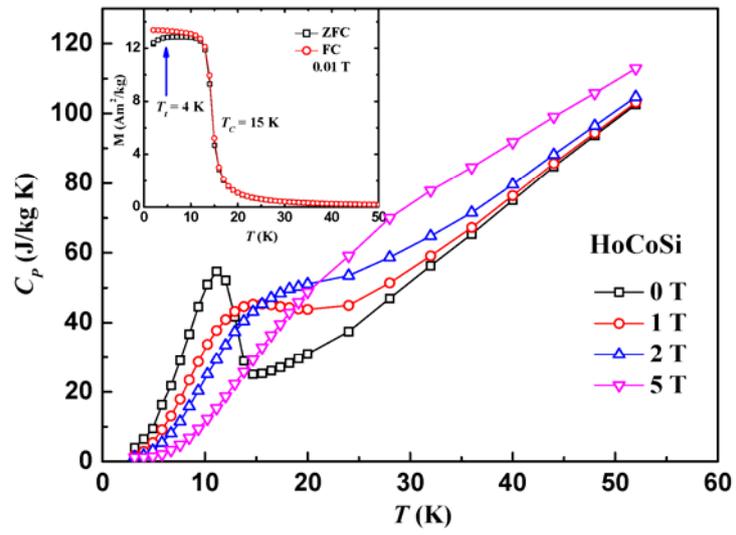

Fig. 5

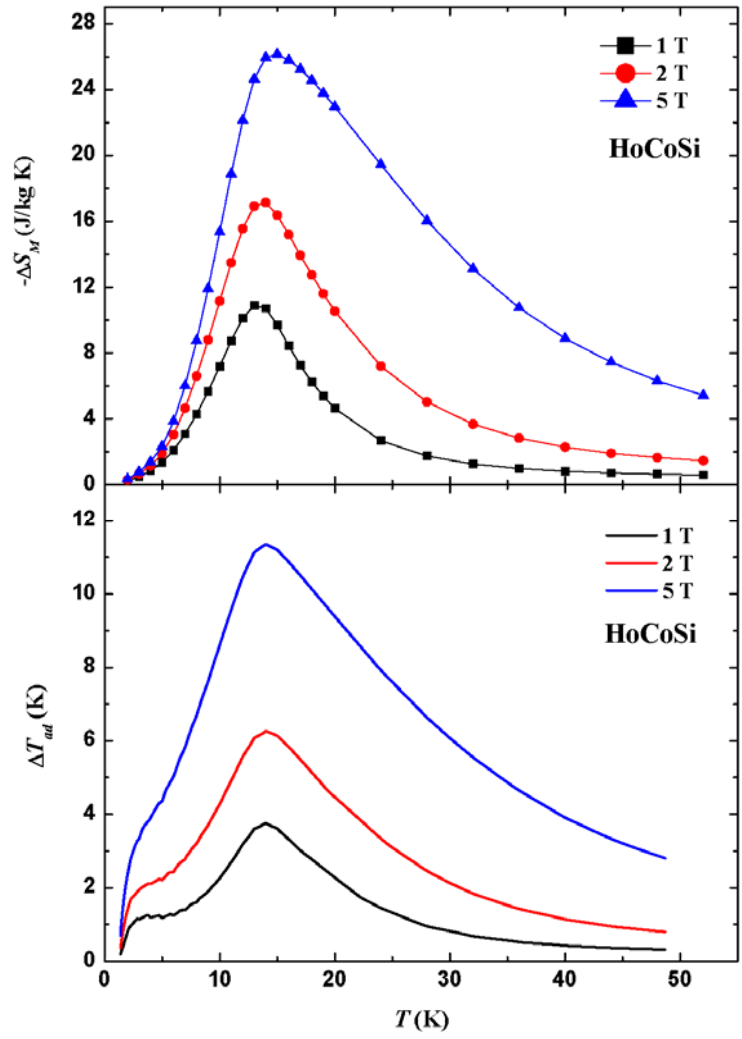

Fig. 6

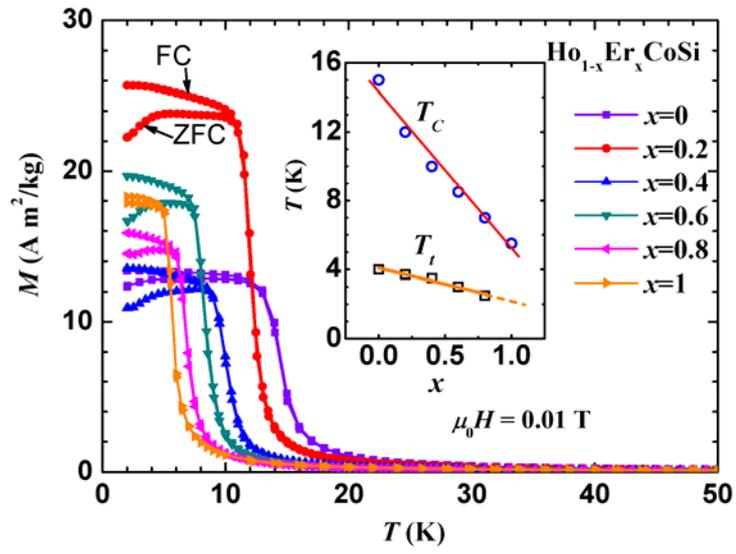

Fig. 7

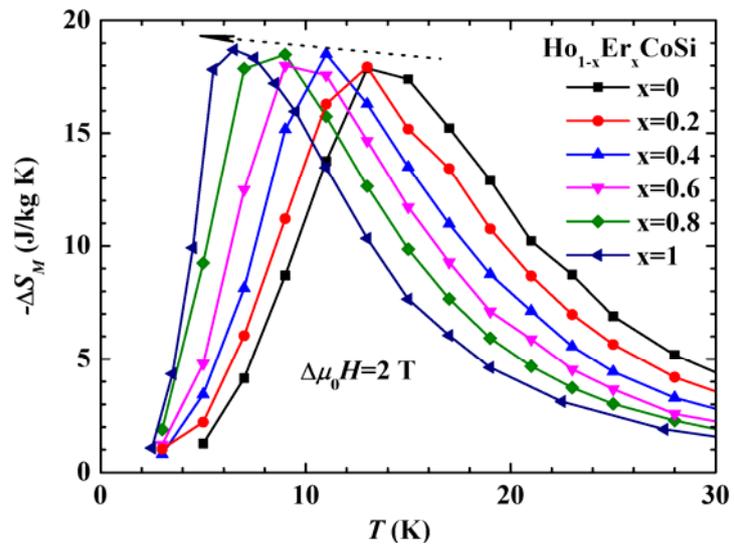

Fig. 8

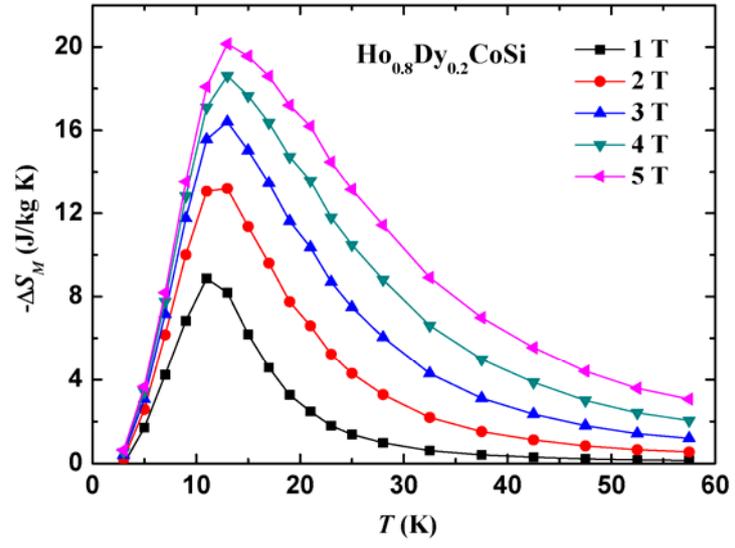

Fig. 9

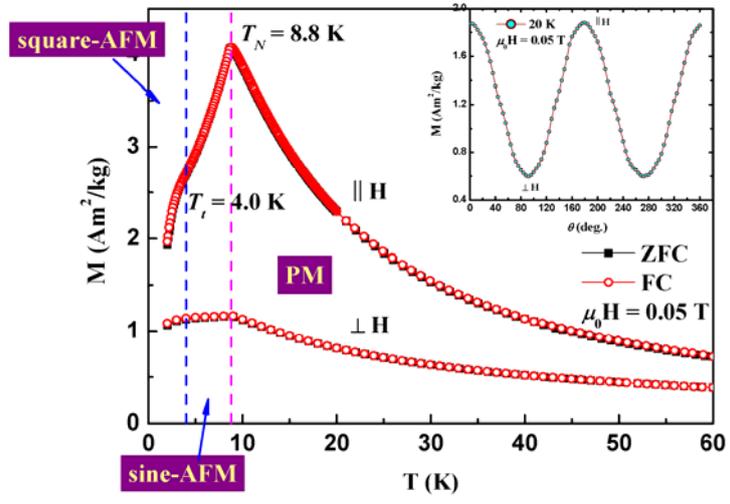

Fig. 10

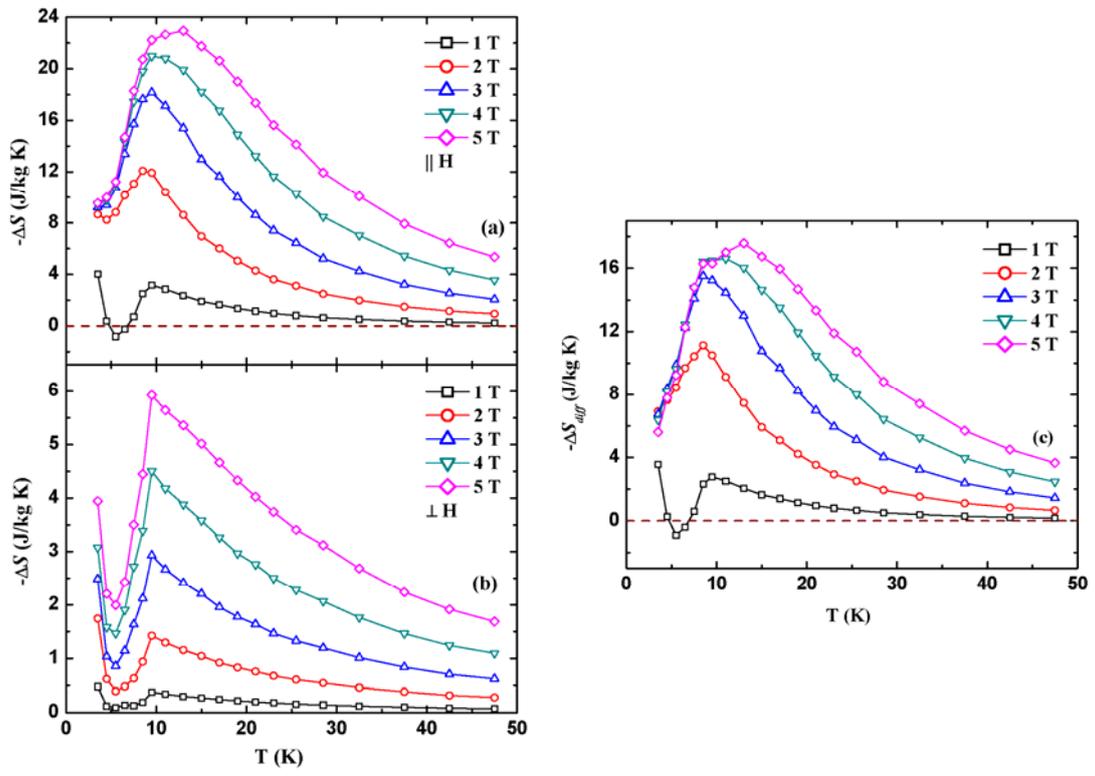

Fig. 11

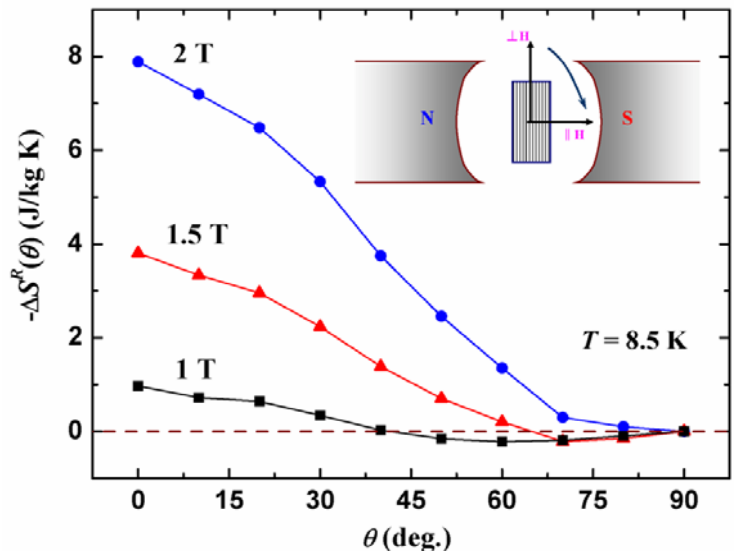

Fig. 12

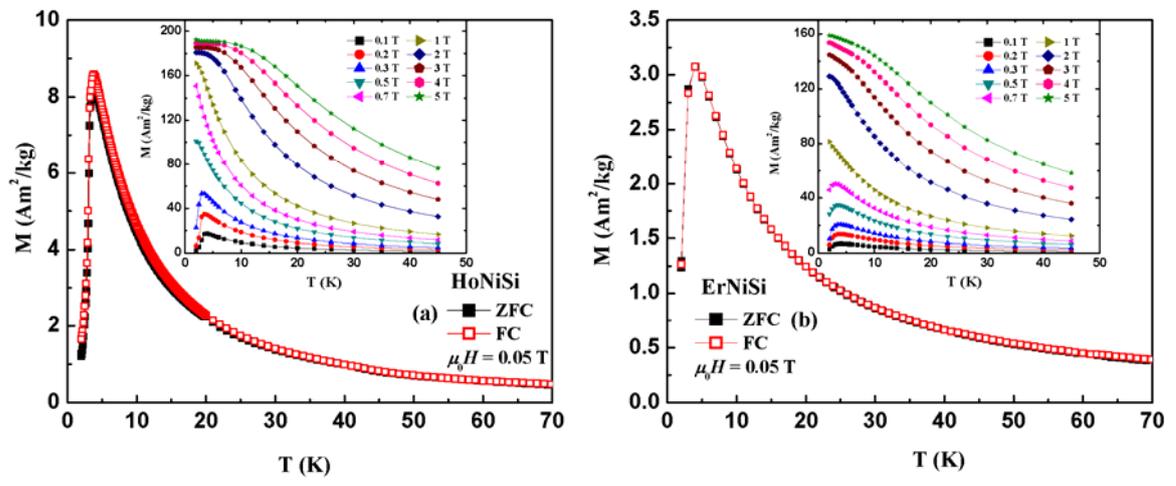

Fig. 13

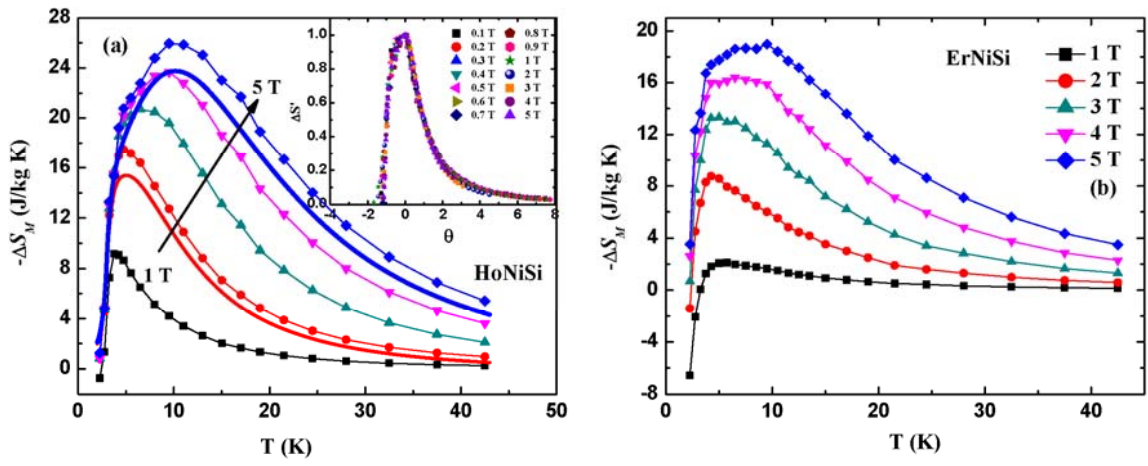

Fig. 14

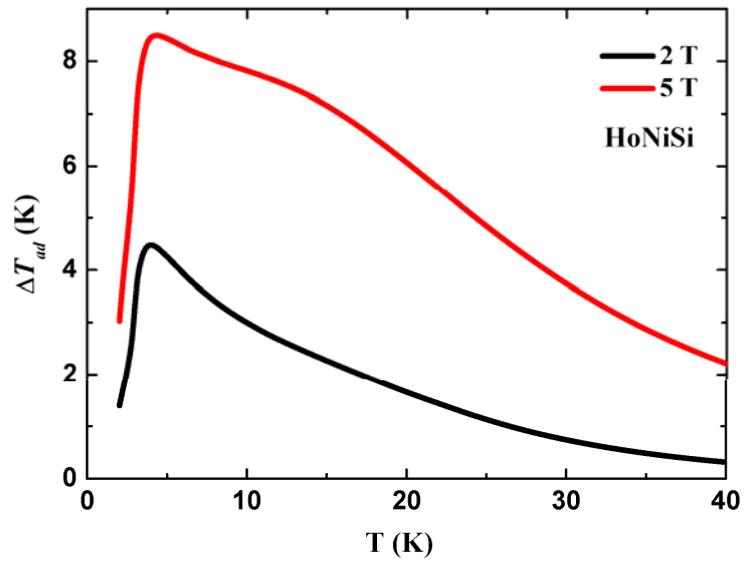

Fig. 15

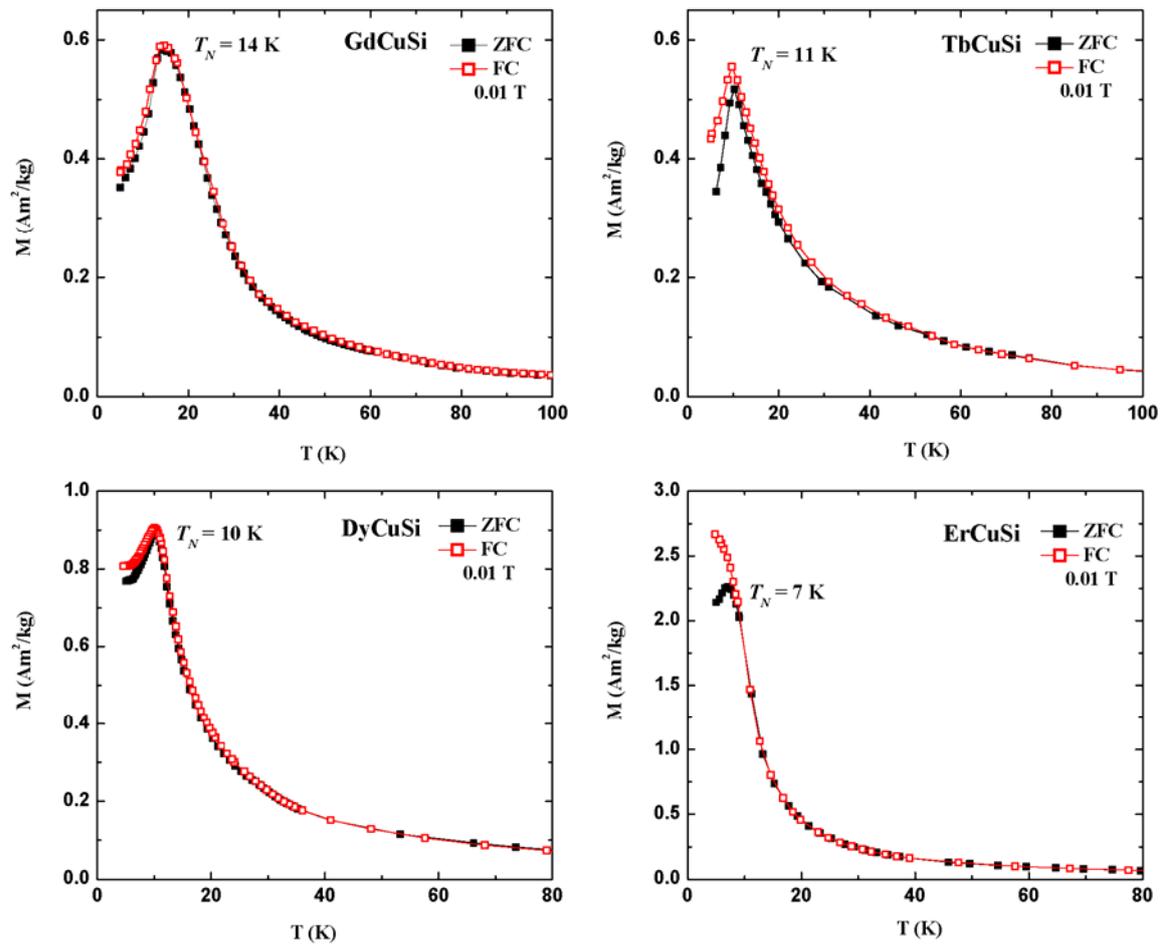

Fig. 16

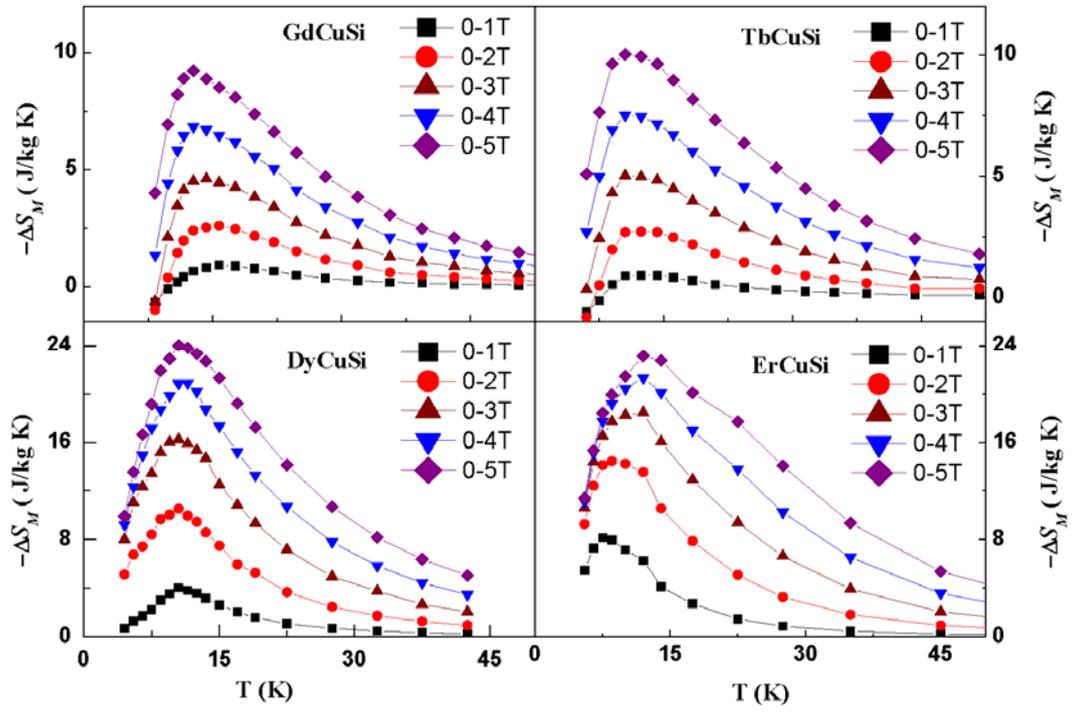

Fig. 17

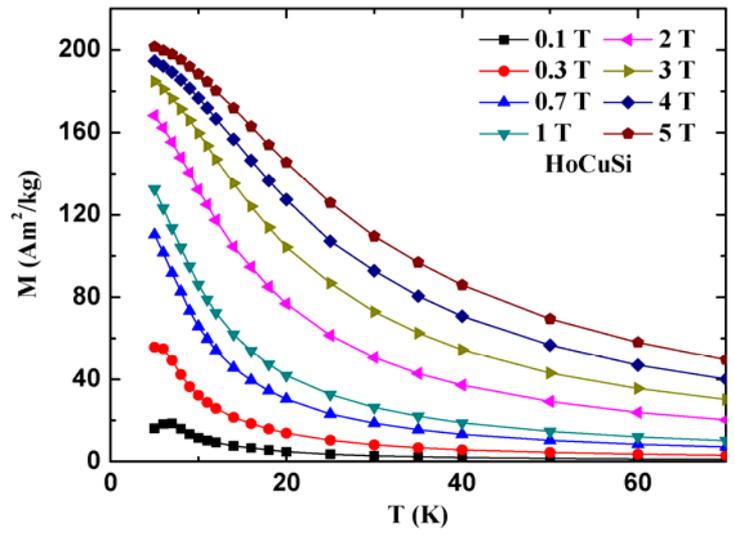

Fig. 18

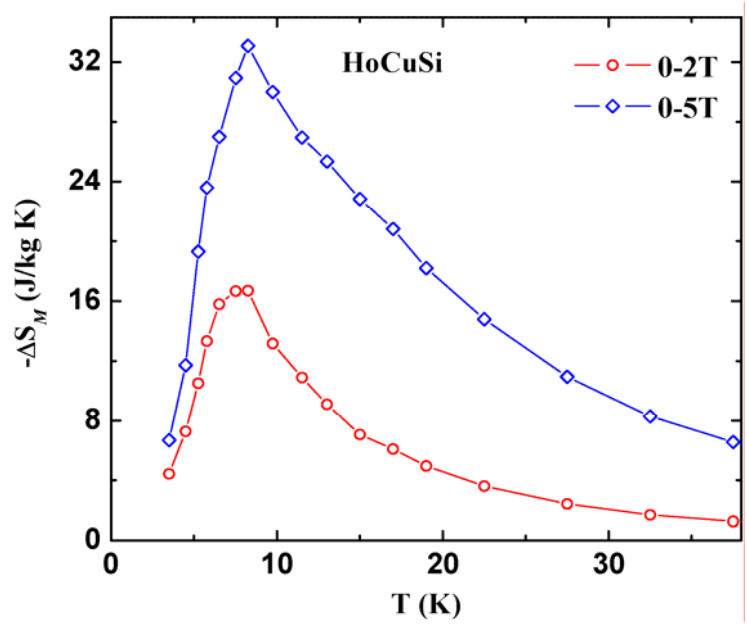

Fig. 19

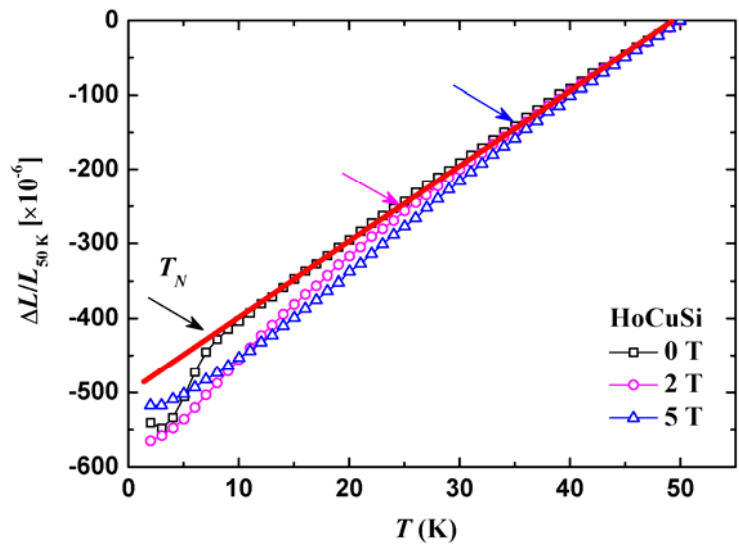

Fig. 20

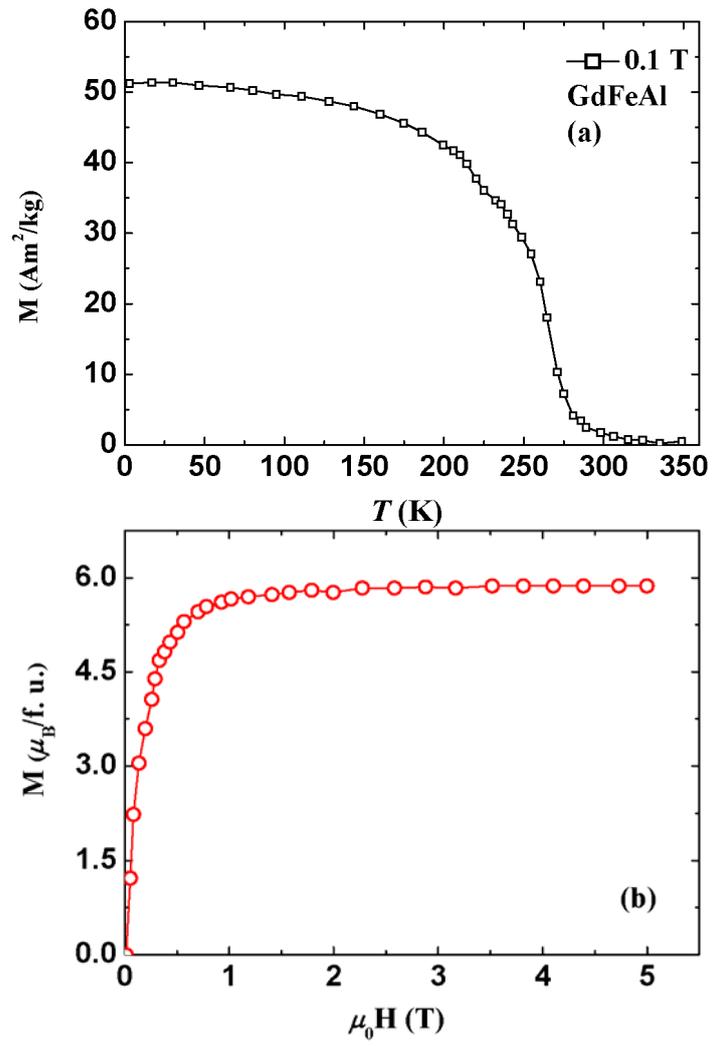

Fig. 21

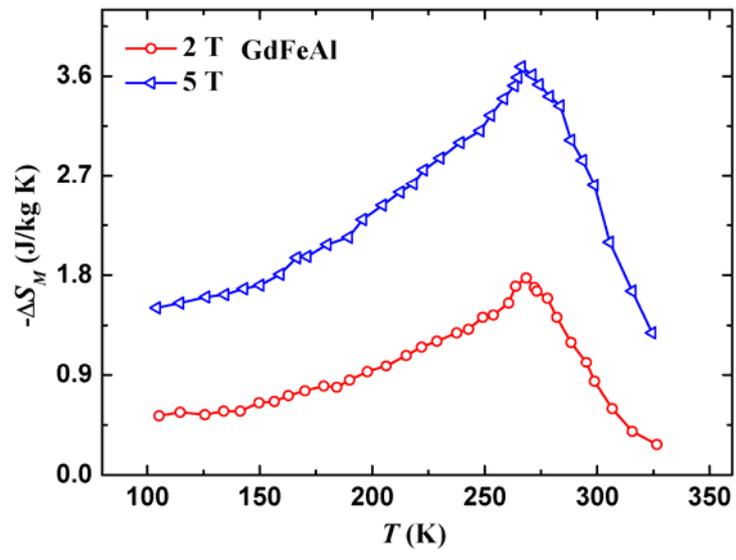

Fig. 22

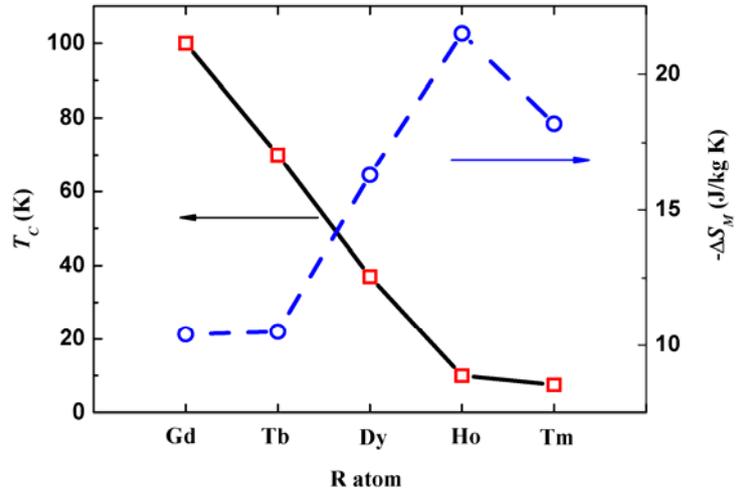

Fig. 23

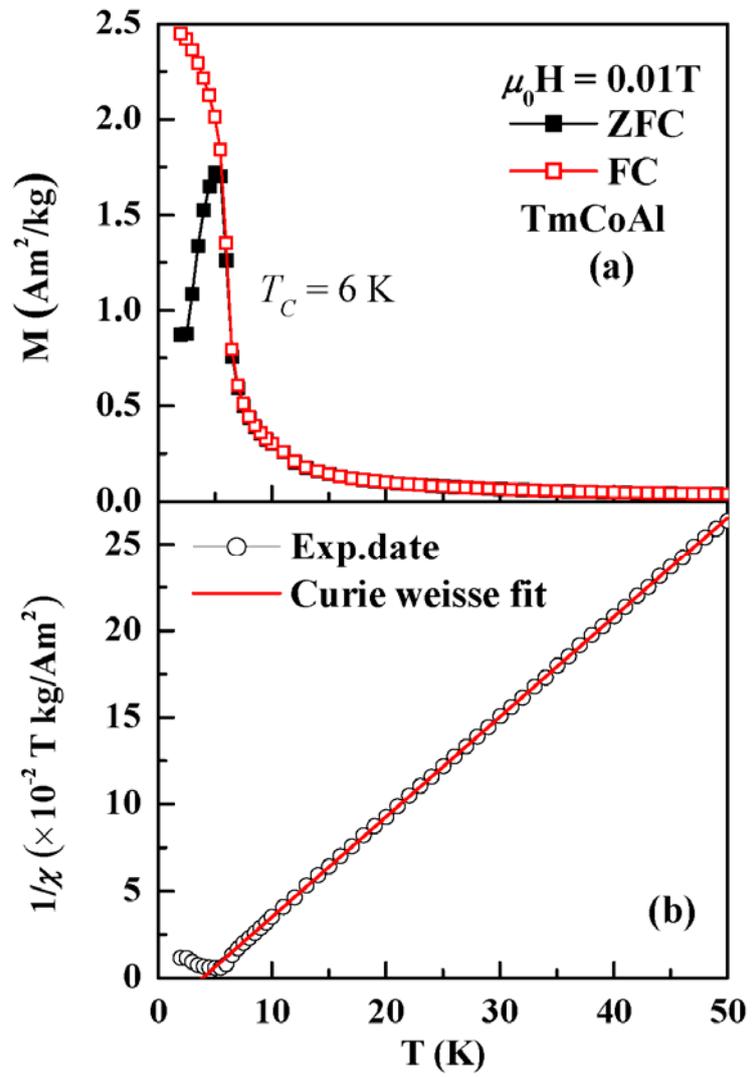

Fig. 24

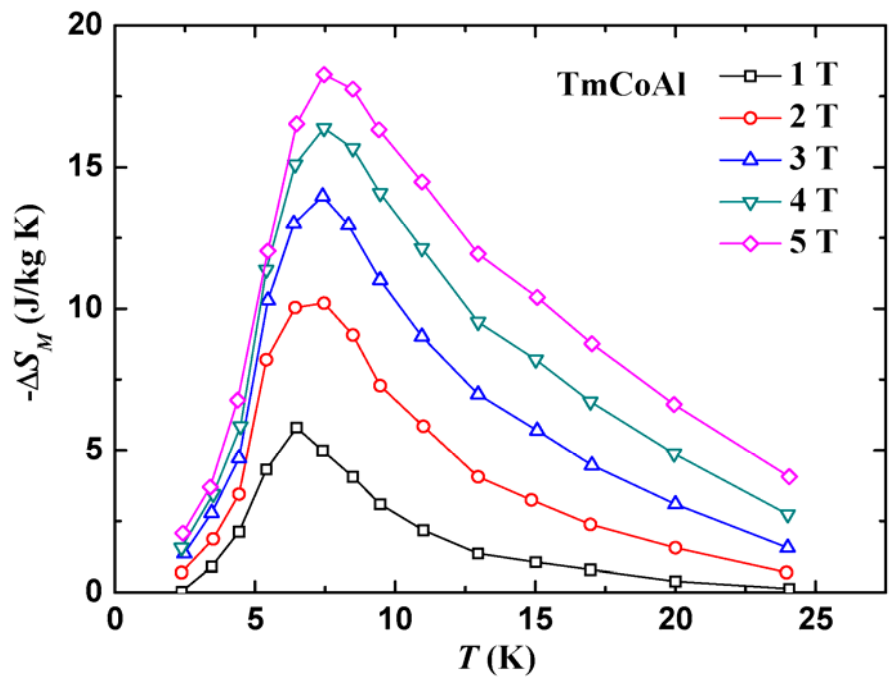

Fig. 25

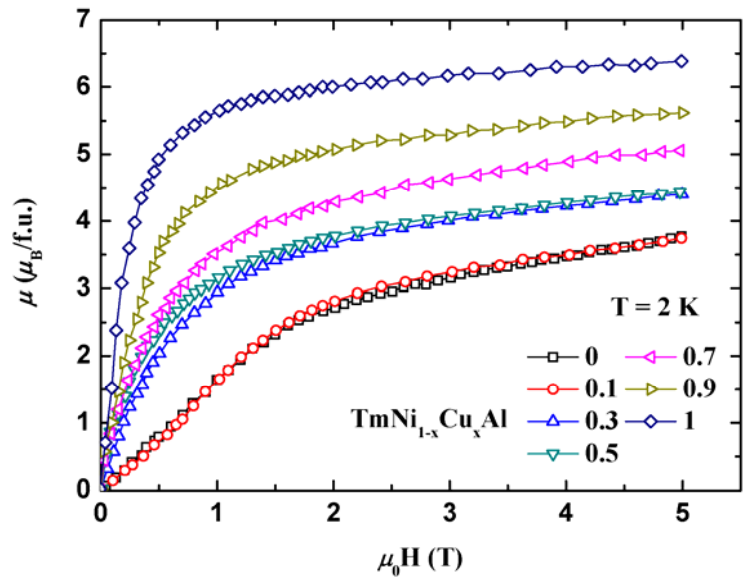

Fig. 26

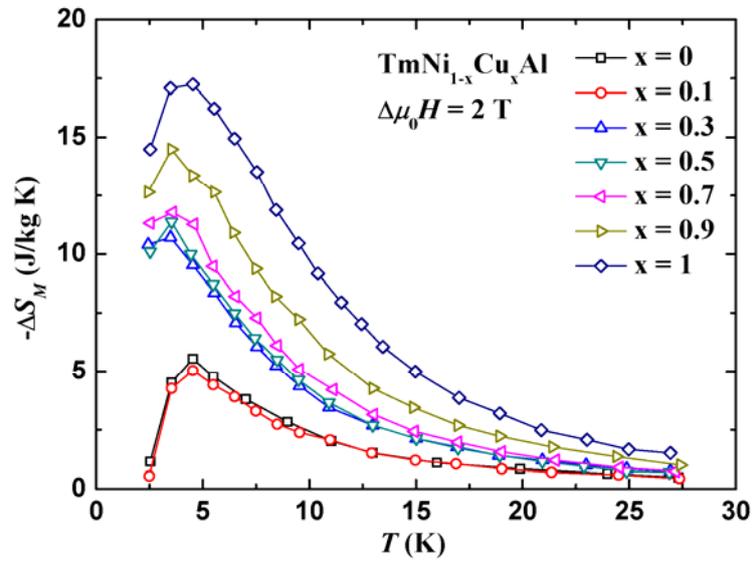

Fig. 27

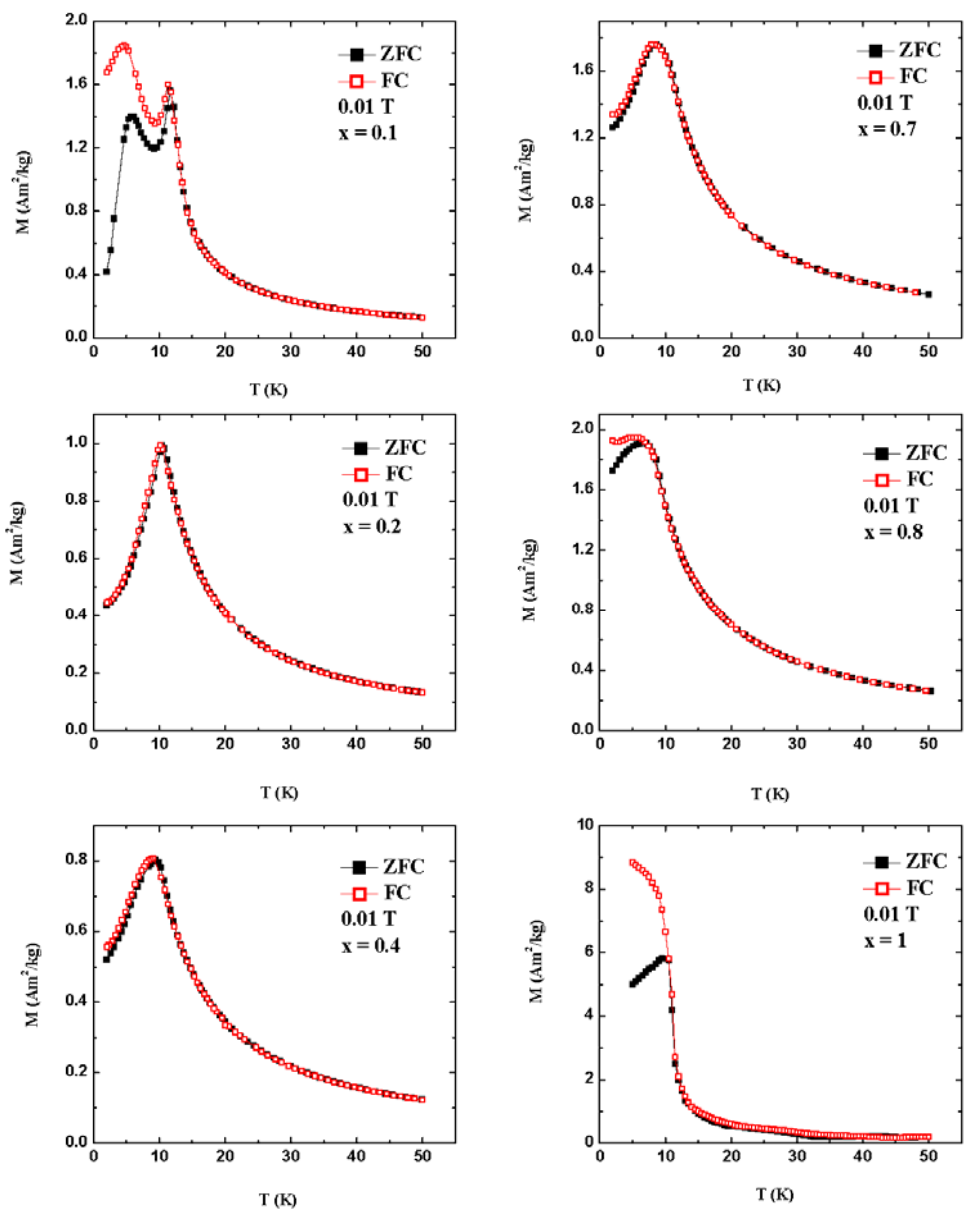

Fig. 28

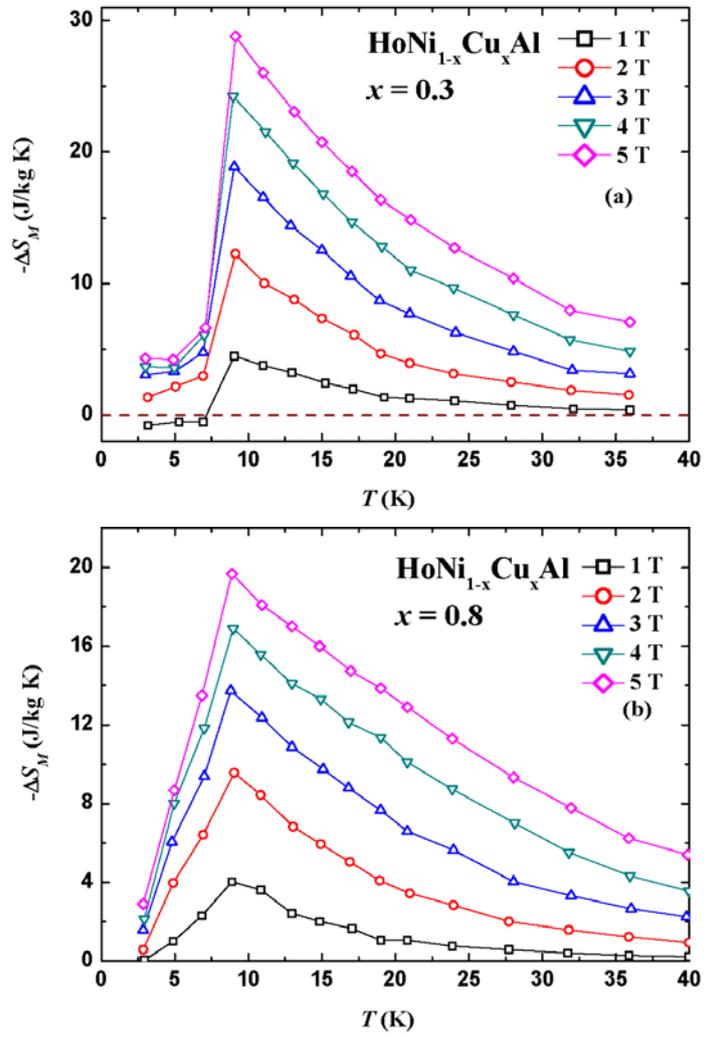

Fig. 29

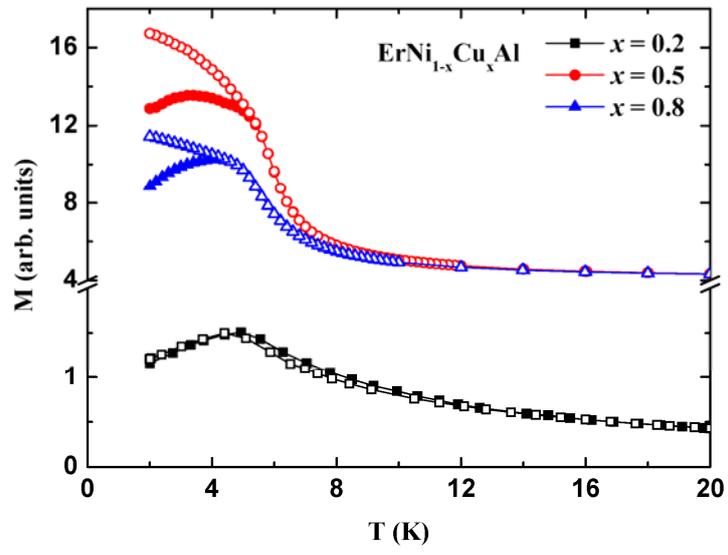

Fig. 30

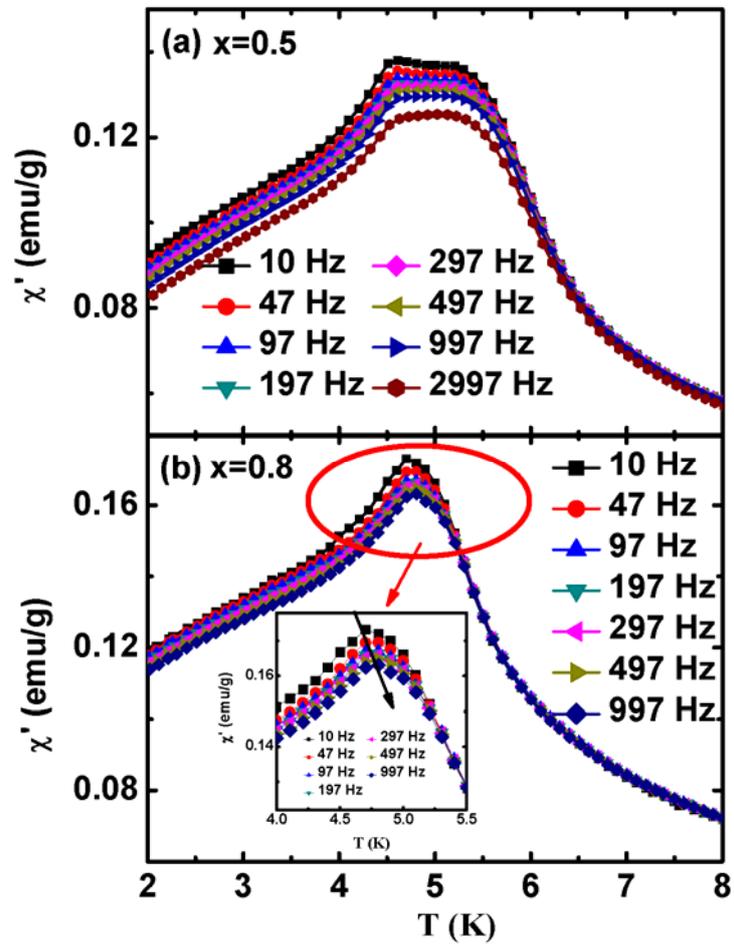

Fig. 31

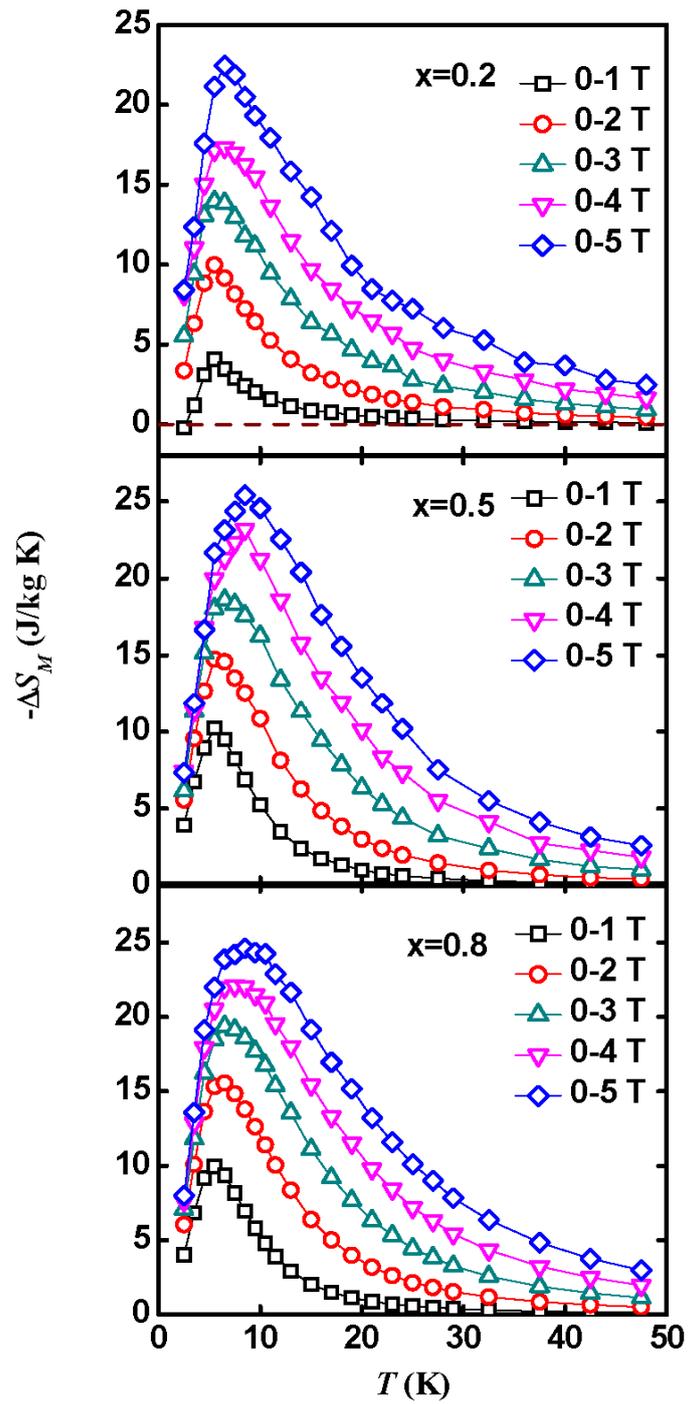

Fig. 32

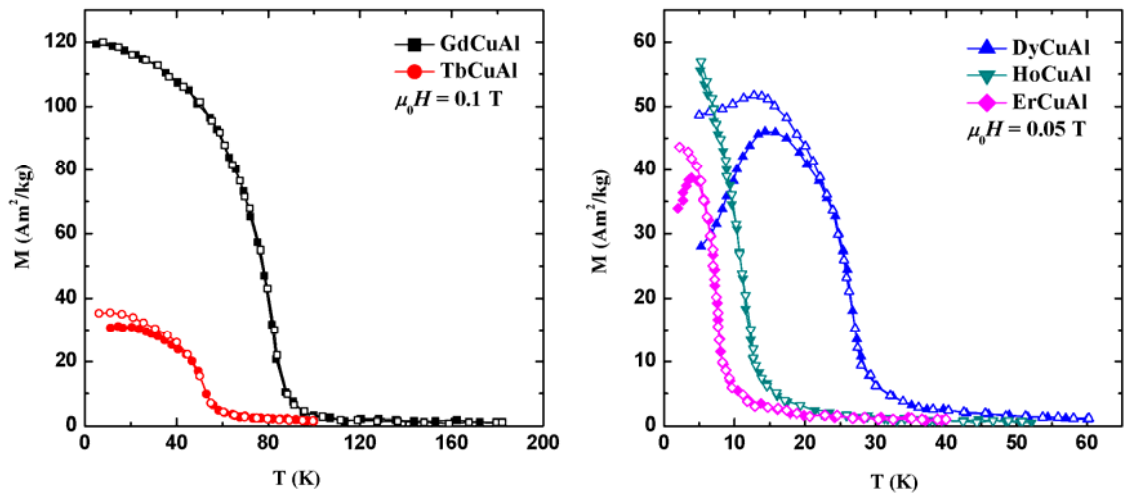

Fig. 33

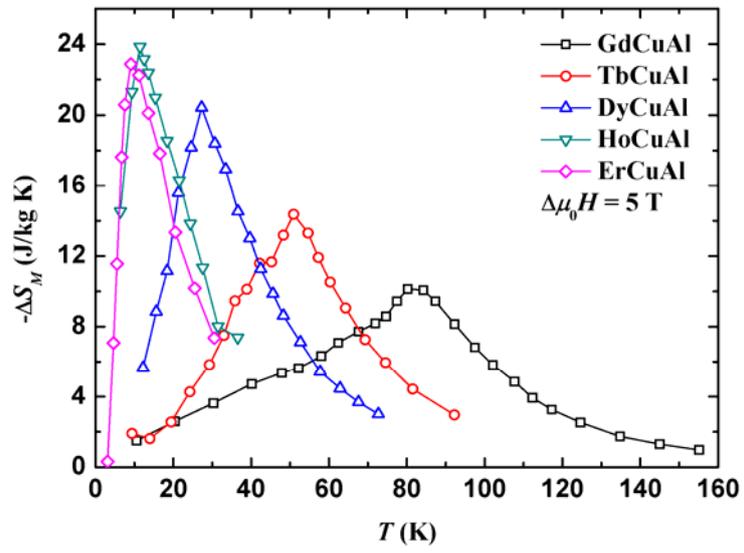

Fig. 34

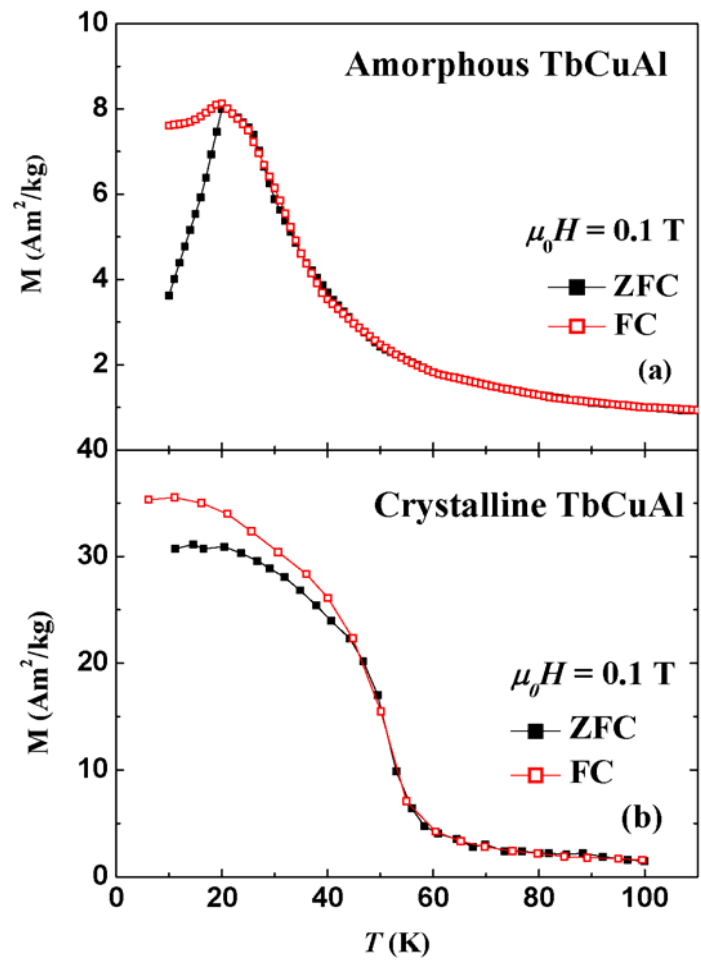

Fig. 35

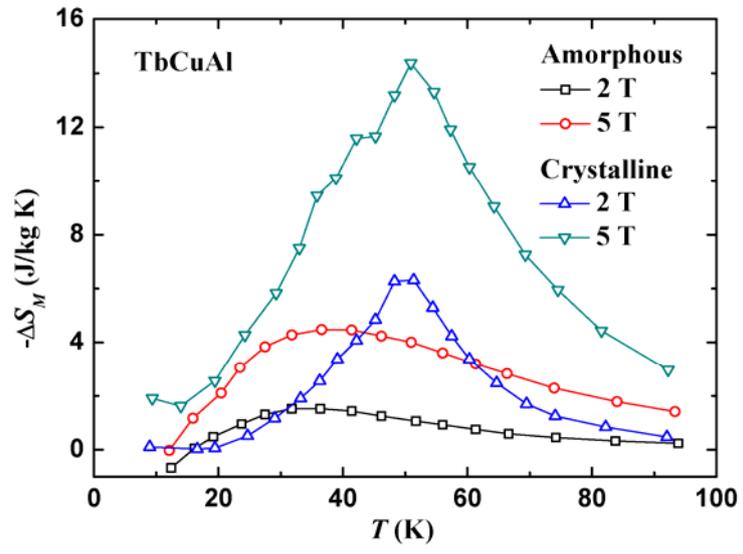

Fig. 36

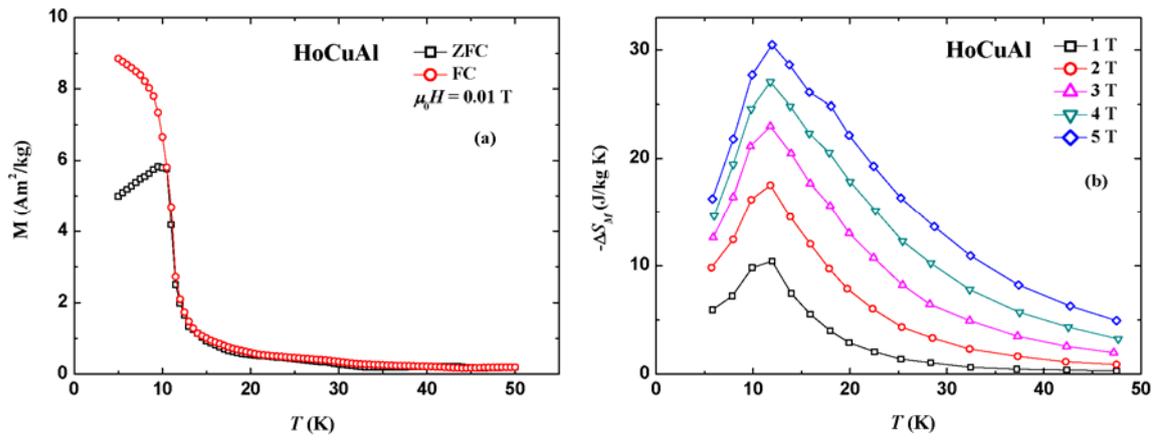

Fig. 37

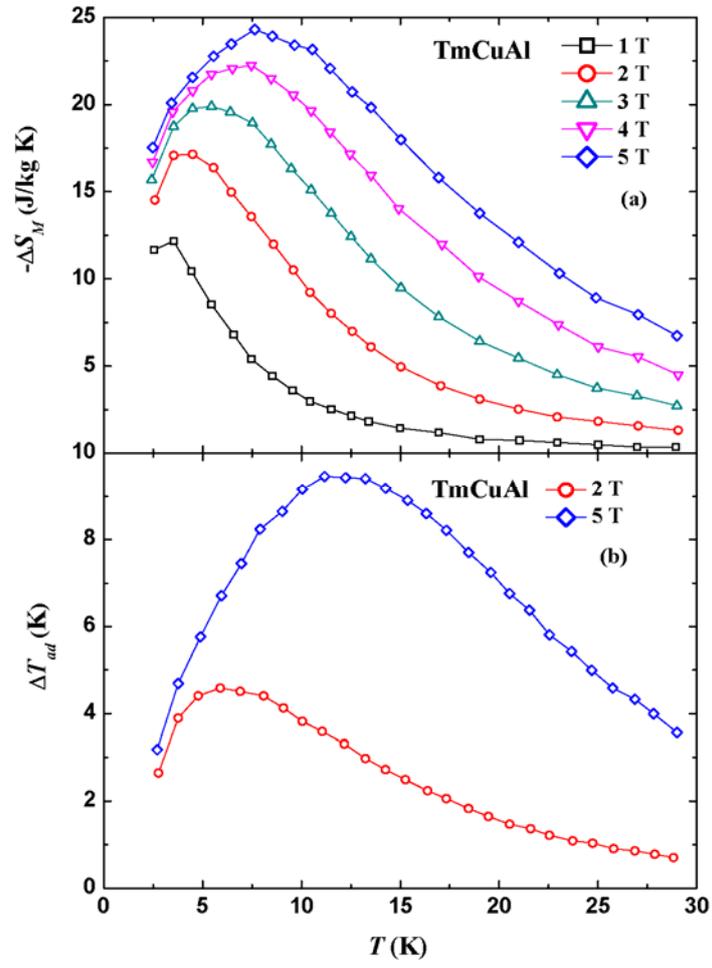

Fig. 38

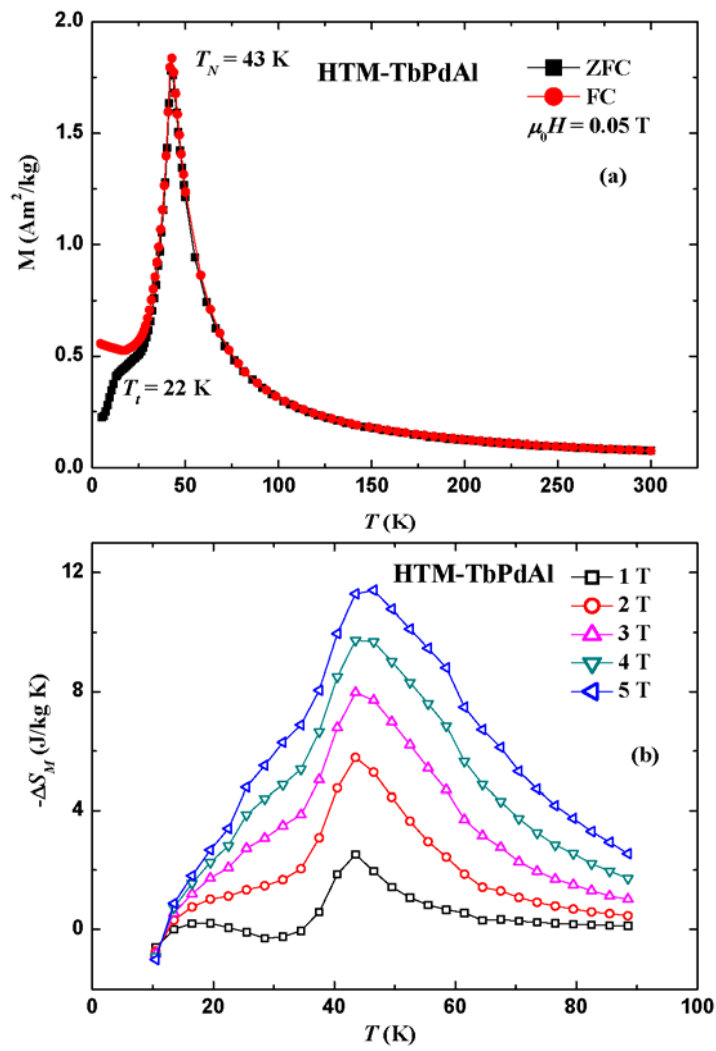

Fig. 39

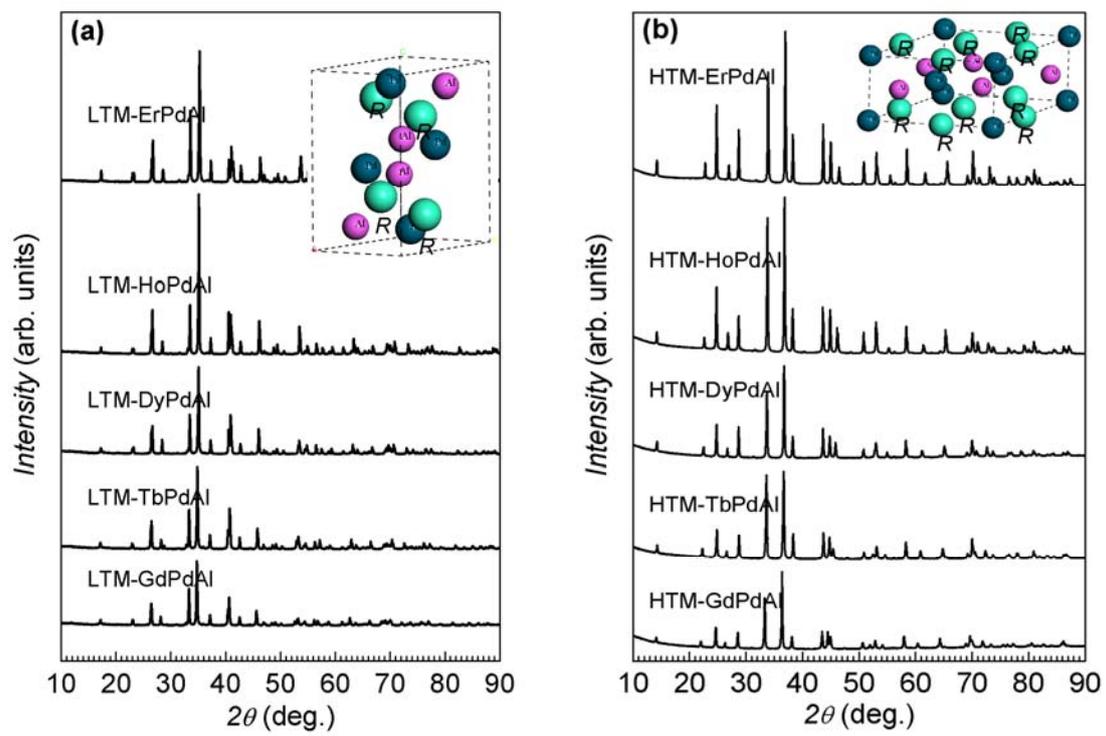

Fig. 40

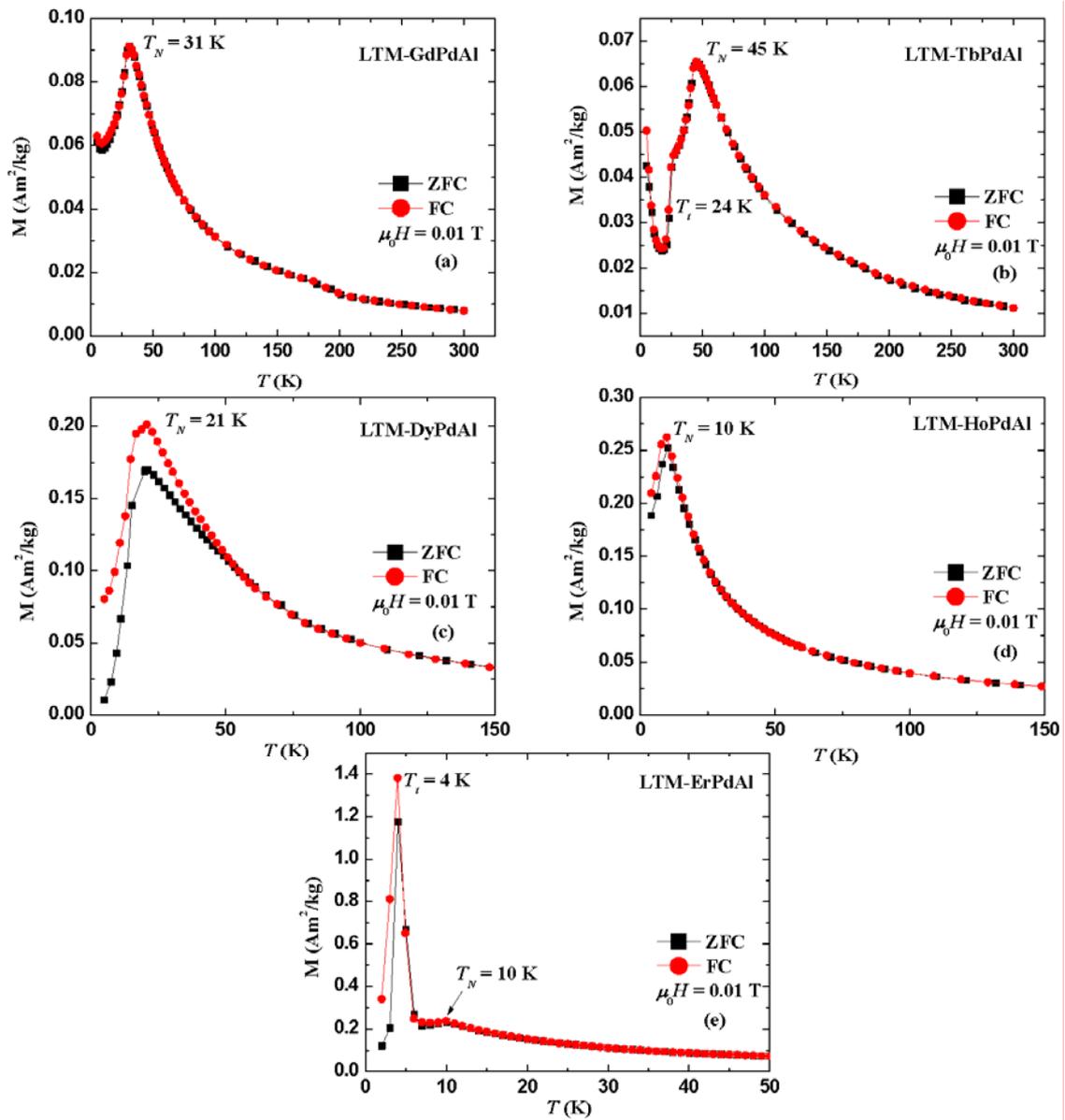

Fig. 41

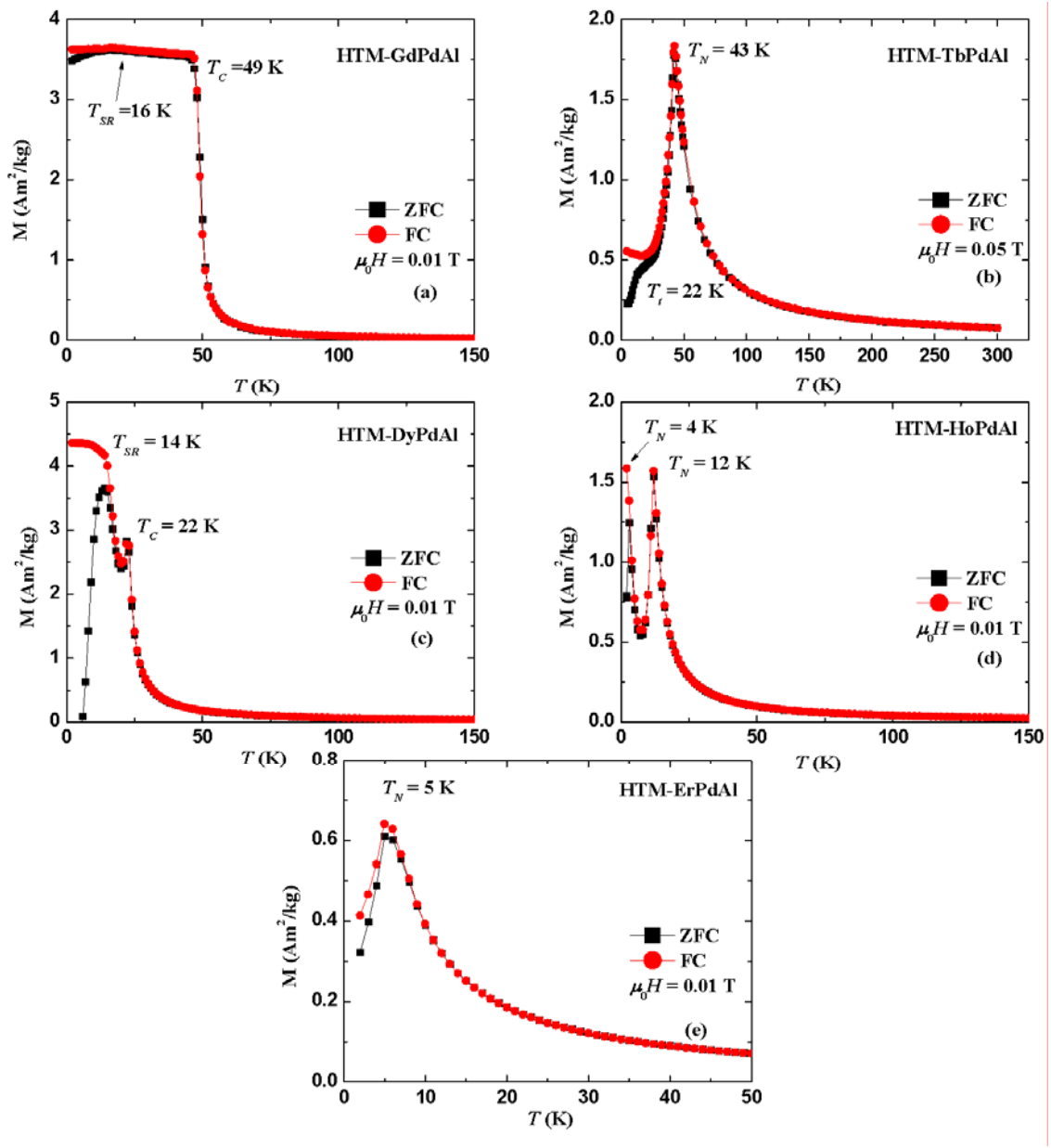

Fig. 42

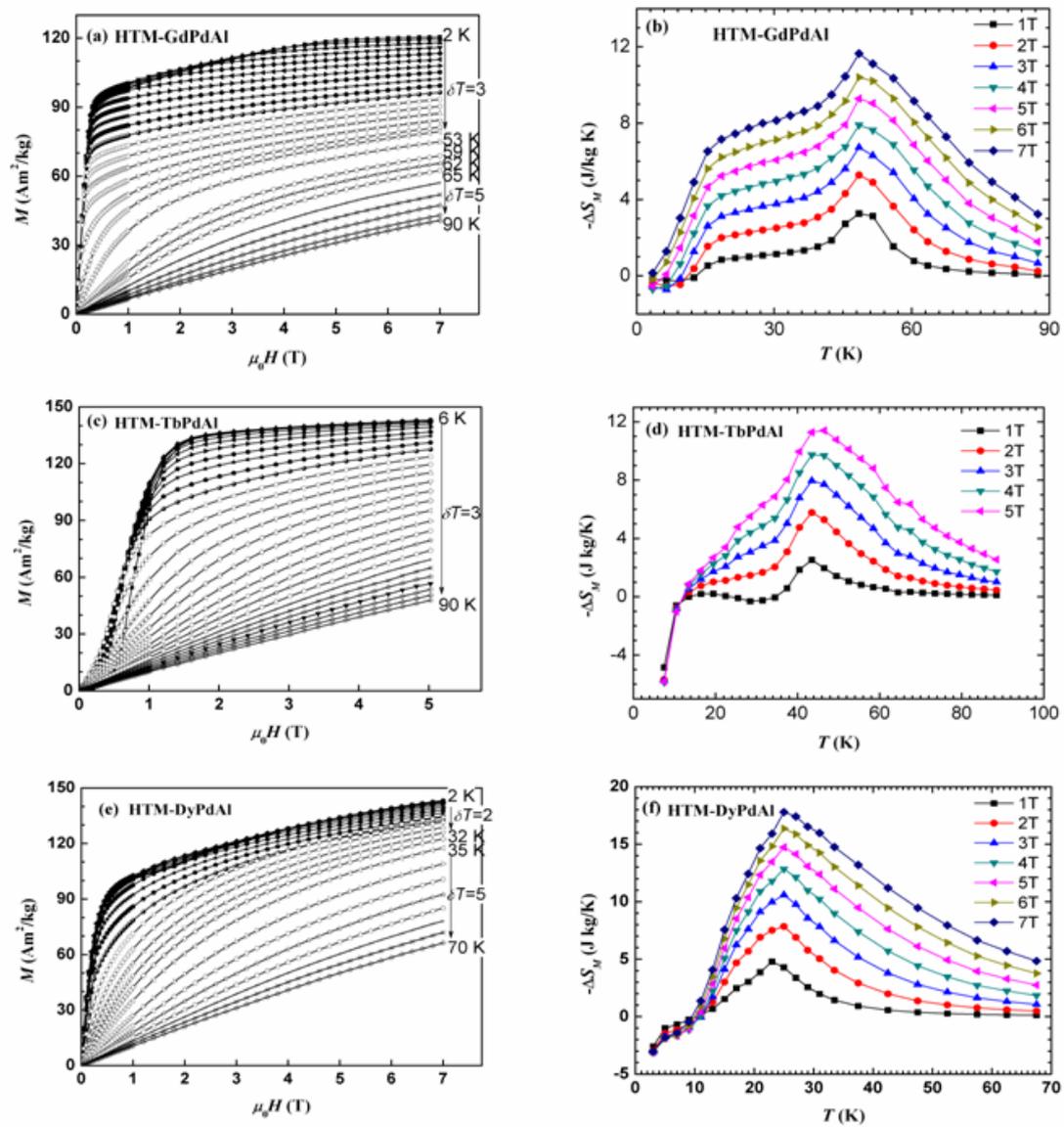

Fig. 43

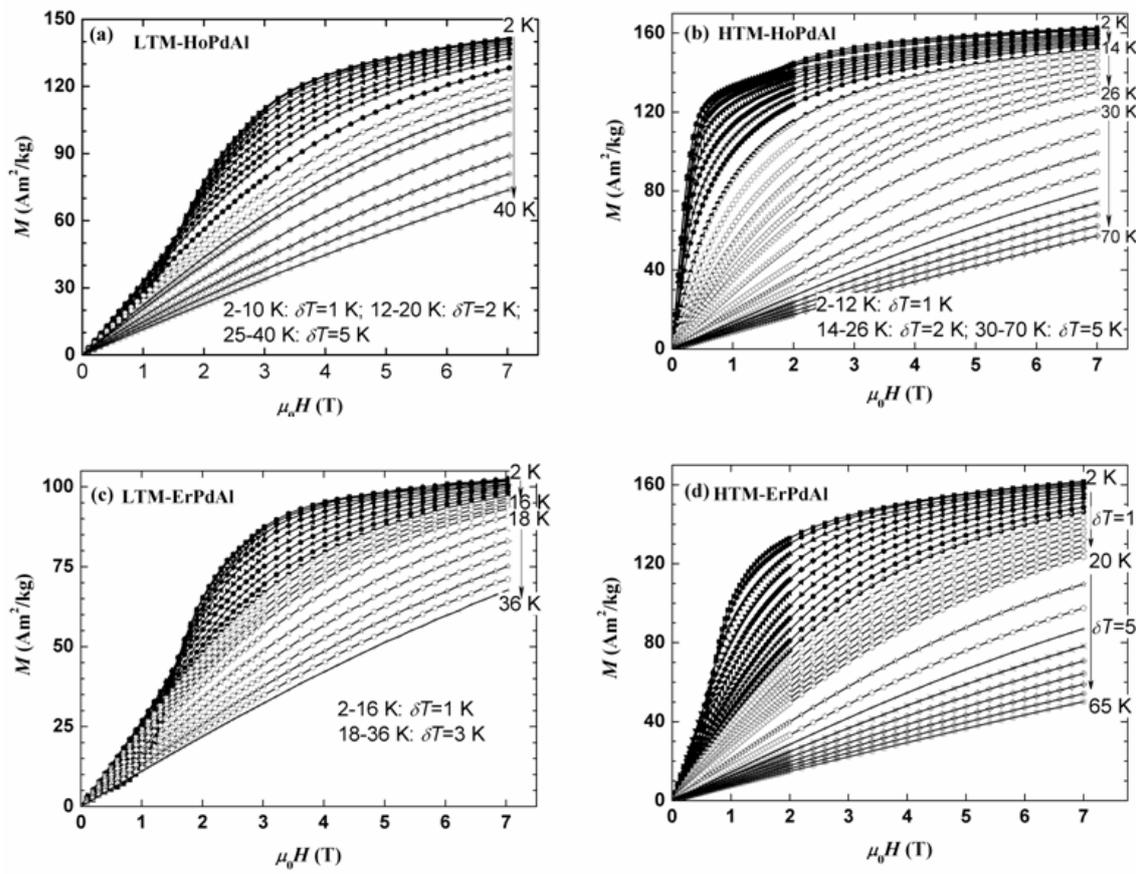

Fig. 44

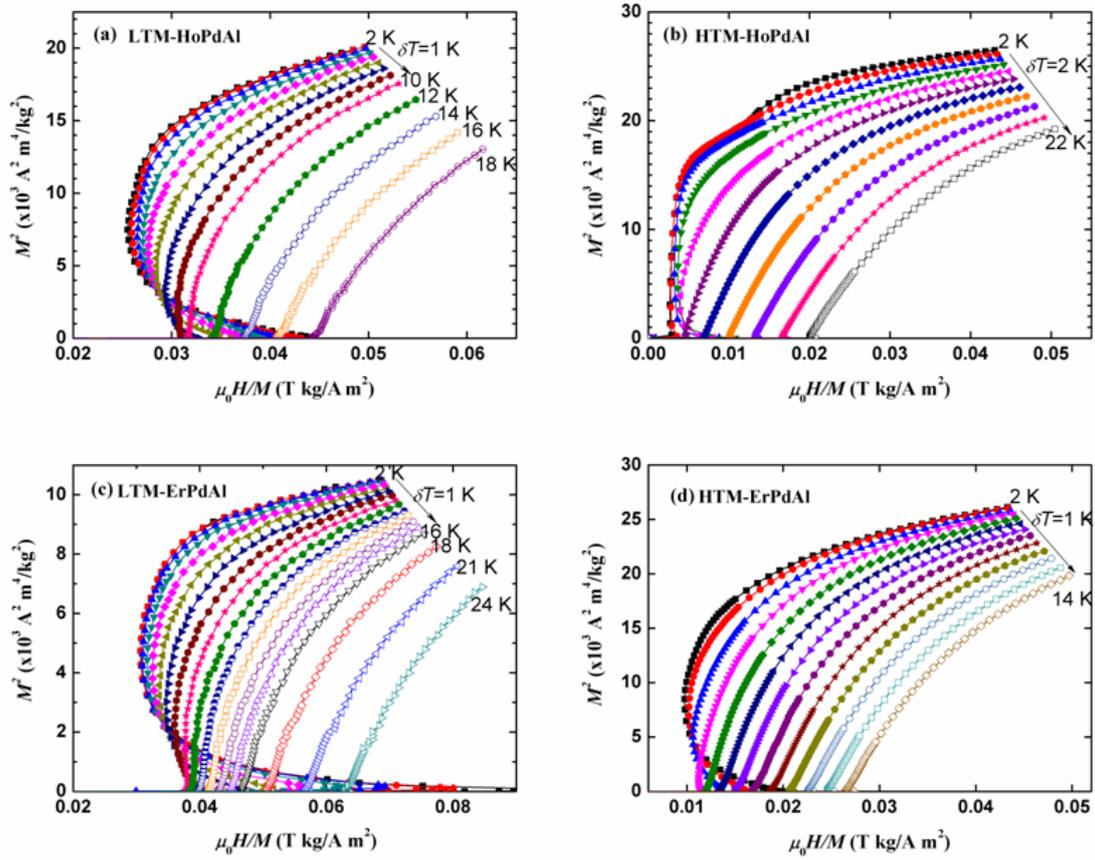

Fig. 45

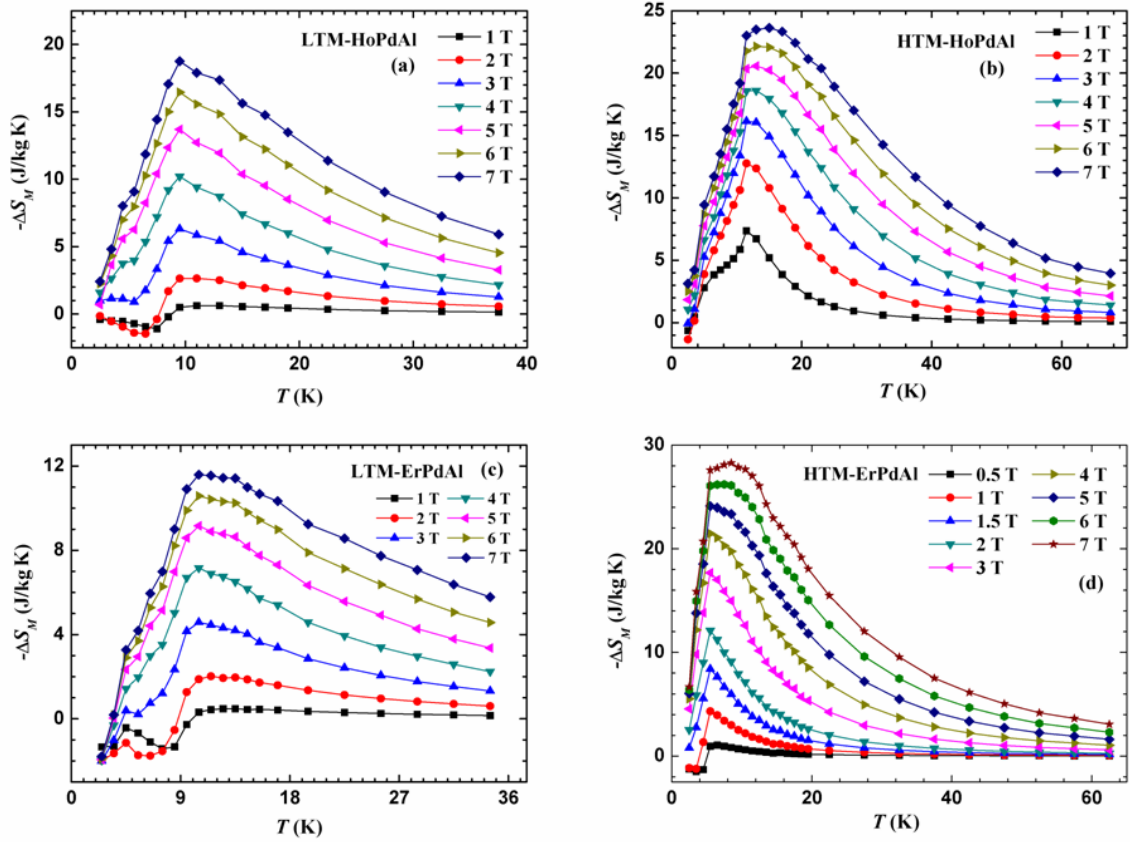

Fig. 46